\documentclass[prd,
               twocolumn,
               eqsecnum,
               showpacs,
               letterpaper,
               superscriptaddress,
               altaffilletter,
               nofootinbib, nobibnotes
               ]{revtex4-1}
 \usepackage[utf8x]{inputenc}
 \usepackage{graphicx}
  \usepackage{amsmath,amssymb} 
 \usepackage[english]{babel}
  \usepackage{rotating}
  \usepackage{amsbsy}
 \usepackage{textcomp}
 \usepackage{psfrag}
 \usepackage{multirow}
 \usepackage[usenames,dvipsnames]{color} 
 \usepackage{sidecap}
 \usepackage{array}
 \newcommand{\beq }{\begin{equation}}
\newcommand{\eeq}{ \end{equation}}
\newcommand{\beqa }{\begin{eqnarray}}
\newcommand{\eeqa }{\end{eqnarray}}
\newcommand{\bwt }{\begin{widetext}}
\newcommand{\ewt }{\end{widetext}}

\newcommand{\Arxiv}{\emph{ArXiv}}
\newcommand{\CQG}{\emph{Class. Quantum Grav. }}
\newcommand{\apjl}{\emph{Astrophys. J. Letters }}
 
\def\col{color}
\def\Figscale{0.3}

\def\SkyFigscale{0.4}

\def\H{\mathcal{H}}
\newcommand{\pvec}{\vec{\theta}}

\begin{document}

\title{Effect of calibration errors on Bayesian parameter estimation for gravitational wave signals from inspiral binary systems in the advanced detectors era}
\author{Salvatore~Vitale}
\author{Walter~Del~Pozzo}
\author{Tjonnie~G.~F. Li}
\author{Chris~Van~Den~Broeck}
\affiliation{Nikhef, Science Park 105, 1098 XG Amsterdam, The Netherlands}
\author{Ilya Mandel}
\author{Ben Aylott}
\affiliation{School of Physics and Astronomy, University of Birmingham,  Edgbaston, Birmingham B15 2TT, UK}
\author{John Veitch}
\affiliation{School of Physics and Astronomy, Cardiff University, Cardiff CF24 3AA, UK}
\begin{abstract}

By 2015 the advanced versions of the gravitational-wave detectors Virgo and LIGO will be online. They will collect data in coincidence with enough sensitivity to potentially deliver multiple detections of gravitation waves from inspirals of compact-object binaries.  This work is focused on understanding the effects introduced by uncertainties in the calibration of the interferometers. We consider plausible calibration errors based on estimates obtained during LIGO's fifth and Virgo's third science runs, which include frequency-dependent amplitude errors of $\sim 10\%$ and frequency-dependent phase errors of $\sim 3$ degrees in each instrument.  We quantify the consequences of such errors estimating the parameters of inspiraling binaries.  We find that the systematics introduced by calibration errors on the inferred values of the chirp mass and mass ratio are smaller than $20\%$ of the statistical measurement uncertainties in parameter estimation for $90\%$ of signals in our mock catalog.  Meanwhile, the calibration-induced systematics in the inferred sky location of the signal are smaller than $\sim 50\%$ of the statistical uncertainty. We thus conclude that calibration-induced errors at this level are not a significant detriment to accurate parameter estimation.

\end{abstract}

\maketitle 
\section{Introduction}

The detection of gravitational waves (GWs) will give us empirical access to the genuinely strong-field dynamics of space-time and allow us to probe astrophysical phenomena inaccessible through electromagnetic observations alone. Despite indirect proofs, like the shrinking of the orbit in the Hulse-Taylor binary, which is in excellent agreement with the theoretical calculation \cite{HulseTaylor}, a direct detection of GWs is yet to occur.
Gravitational-wave detectors based on interferometry: the two LIGO instruments \cite{ILigo}, VIRGO \cite{IVirgo, IVirgo2} and GEO600 \cite{Geo, Geo2}, have collected data in coincidence trough October 2010.  The most recent published results~\cite{S5_lowM,S5_highM}, which cover the period 4 November 2005 – 30 September 2007, do not claim detections. 
The LIGO instruments and Virgo will undergo major improvements in the next few years, and will begin collecting data again by 2015, with an improved sensitivity \cite{ALigo, AVirgo} that may allow for frequent detections \cite{RatesPaper}, ushering in the so-called advanced detector era.

Apart from the intrinsic scientific importance of a first direct detection, the advanced versions of the instruments will open a new era of astronomy and cosmology, in which GWs will be used to test the strong-field regime of General Relativity \cite{Will:2006,LiPreparation,MishraEtAl:2010,DelPozzoEtAl:2011,CornishEtAl:2011}; to set better bounds for the values of the cosmological parameters \cite{StandardSirens,DalalEtAl:2006,MacLeodHogan:2008,Schutz2009,SathyaEtAl:2010,NissankeEtAl:2010,ZhaoEtAl:2011,DelPozzo,TaylorEtAl:2011}; to check the validity of the equations of state for neutron stars \cite{EOS}; to probe the astrophysics of binary evolution \cite{MandelOShaughnessy:2010}; etc.

In order to extract as much physical information as possible, all the known sources of error must be eliminated, reduced or quantified.
Among the known sources of errors, there are \emph{calibration errors}, i.e.~errors on the measurement of the transfer function, which converts the readout of the instruments to the strain used for data analysis. 

These errors will have consequences for the estimation of the intrinsic and extrinsic parameters of the source of GWs, as the data analyst will infer an incorrect data stream. 
Some previous works have dealt with calibration errors, in the context of detection efficiency using template banks \cite{Lindblom2008} and parameter estimation \cite{Bose2005}, but a complete treatment requires the use of numerical methods, because the high dimensionality of the problem and the correlations between the unknown parameters on which the GWs depend make it impossible to forecast the exact effects of calibration errors analytically. 

In this article we have used a Bayesian approach to study and quantify these effects for the first time in the literature.  We created catalogs of 250 software injections (i.e.~signals of known shape added to synthetic noise) in each of three mass bins: one for binary neutron star systems, one for binary black holes, and one for neutron star-black hole systems. We have generated ten different sets of calibration error curves, with shapes and magnitudes that should be representative of the errors we expect to have in the advanced detector era.

The catalogs of injections were analyzed twice: first, by running a Bayesian parameter-estimation code \cite{vanderSluysEtAl:2008,Veitch2010} on the original injections, and then by running the same code after artificially adding, one at a time, the calibration errors we had generated.
As the presence of the errors was the only thing that had changed between one analysis and the other, the differences observed in the recovered parameters and the Bayes factors could only have been caused by the calibration errors, and we were able to quantify these differences and relate them to the calibration errors.

We have found that the effects are generally small, the shifts introduced in the estimated parameters being a fraction of the statistical measurement errors due to the noise in the instruments. At the same time, the Bayes factors of the signals are only slightly affected by the errors we have considered, the average shift being $\sim 0.9\%$, so that if the Bayes factor were used as a detection statistic, in the way described in~\cite{Veitch2010}, there will not be signals that are going to be missed because of the way the errors have changed their shapes. 

This article is organized as follows:
In Sec. \ref{Sec.CalibrationTechniques} we describe the interferometers and the process of calibration.\\
In Sec. \ref{Sec.CalibrationErrors} we describe the errors associated with the calibration process, and how we model them.\\
In Sec. \ref{Sec.BayesianPE} we give some details about the Bayesian approach to parameter estimation and model selection, with specific focus on gravitational-wave data analysis.\\
In Sec. \ref{Sec.Method} we describe the method we have used to quantify the effects of calibration errors, and in the next Sec. \ref{Sec.Results} we report the main results of our analysis. \\

\section{Calibration techniques}\label{Sec.CalibrationTechniques}

{
Ground-based laser interferometric gravitational wave detectors operate in a Michelson interferometer type configuration, measuring the phase propagation difference between two perpendicular arms with a phase accuracy of $\lambda/10^{12}$ ($\lambda$ being the wavelength of the laser). In LIGO and Virgo, this is accomplished by enhancing the GW induced phase changes using 4 km long Fabry-Perot resonators in each of the interferometer arms, optimizing the integration time of the detector to GWs of a few hundreds of hertz. In order to analyze the effects of calibration errors on parameter estimation, as we seek to do in this article, we abstract the incredibly complex interferometer to a single degree of  freedom sensor, only sensitive to differential arm length (DARM) changes, which are expected to contain the gravitational wave signals.
In order to operate such a sensor in a continuous fashion, the DARM signals are measured in closed loop feedback, correcting the measured deviations and keeping the interferometer at the desired operating point. A reduced block schematic of the feedback loop involved is shown in Fig.~\ref{Fig.OLG}.

\begin{figure}[h]
\includegraphics{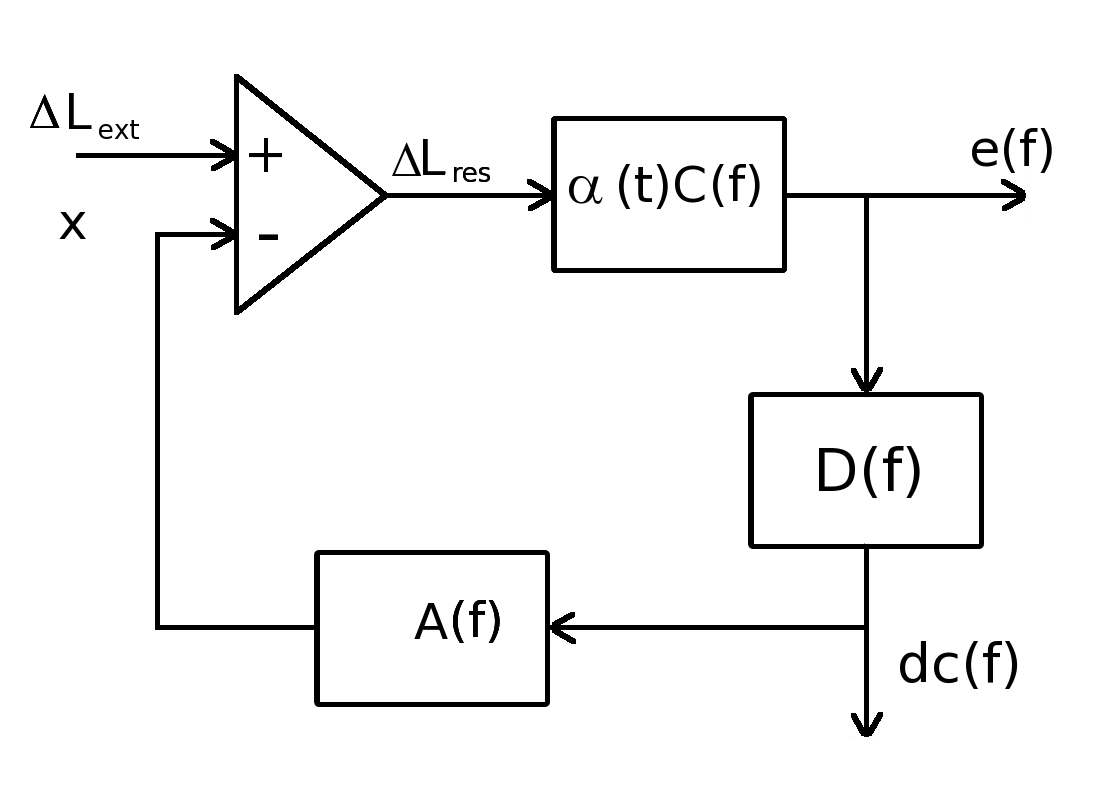}
\caption{A schematic representation of the IFO with the subsystems described in the text.}\label{Fig.OLG}
\end{figure}

The schematic immediately indicates some kind of `in-loop' measurement, where any disturbance is suppressed by the control loop, leaving the interferometer output dependent on the performance of the feedback. In order to reconstruct the actual GW signal, we require accurate knowledge (transfer functions) of all components within the feedback loop. It is the uncertainty in the overall loop transfer function that provides us with an error on the calibration of our gravitational wave detector.
The sensing method used provides the differential phase measurement at the output of the interferometer and is based on the Pound-Drever-Hall (PDH) technique \cite{Drever1983,Regehr1995}, Within the necessary bandwidth, the PDH technique provides a signal, $e(f)$, also called the error signal, that is proportional to the measured deviation. With reference to Fig.~\ref{Fig.OLG}, we see that the external length perturbations, $\Delta L_{ext}$, transfer to the error signal by
\begin{equation}
\Delta L_{ext}(f)=R(t,f)e(f),
\label{ErrorToDarm}
\end{equation}
where, $e(f)$, is the error signal output coming from the interferometer and $R(t,f)$ is the frequency dependent response of the closed loop feedback control system (the time dependence being there to recall that the behavior of the instrument changes with the time, see below).  Within the interferometer calibration nomenclature, $R(t,f)$ is usually referred to as the length response function and completely describes the transfer function between the residual change in DARM and the digital error signal. The calibration of gravitational wave detectors is an entire study unto itself and much is involved in extracting an accurate response for different components within the feedback loop. Evaluating the blocks in Fig.~\ref{Fig.OLG} shows that calibration of the detector output involves three main subsystems. The uncertainty in each of the subsystem's transfer functions carries with it a source of calibration errors, which defined by
\begin{itemize}
\item The transfer function of the arm cavity $C'(t,f)$, which is also known as the sensing function and can be split into a complex frequency dependent part and a slow varying time dependent part: $C'(t,f)=C(f)\alpha(t)$.
\item The digital filter $D(f)$ is applied to the measured error signal and `shapes' the feedback loop response time and the amount of disturbance rejection from external noise.      
\item The actuation function $A(f)$ transfers the `knowledge' of the filtered error signal into a physical correction force on the interferometer. This can be, for example, the force exerted by a voice coil onto the test masses in the interferometer arms.
\end{itemize}

We can set up a set of self consistent equations that describes the behavior of the closed loop system. With reference to the variables in Fig.~\ref{Fig.OLG} these are,
\begin{eqnarray}
\Delta L_{res}&=&\Delta L_{ext}-x\\ 
\label{eq:2.2}
e(f)&=&\alpha(t)C(f)\Delta L_{res}\\
\label{eq:2.3}
dc(f)&=&e(f)D(f)\\
\label{eq:2.4}
x&=&A(f)dc(f).
\label{eq:2.5}
\end{eqnarray}
Rearranging the equations in Eq.(\ref{eq:2.2}) to Eq.(\ref{eq:2.5}), one can find, after some algebra, the explicit expression for the length transfer function term, $R(t,f)$ as:
\begin{equation}
R(t,f)=\frac{1+\alpha(t)G(f)}{\alpha(t) C(f)}
\label{eq:2.6}
\end{equation}
where we introduced the loop gain function, $G(f)=A(f)C(f)D(f)$, also known as the open loop gain of the system. The loop gain $G(f)$ of the feedback system is obtained by breaking the loop at an arbitrary point and multiplying all subsystems by going round the loop once. When analyzing the performance of our gravitational wave sensor it is useful to create a measurement error budget. For the analysis of calibration errors, the error budget describes the noise sources introduced by the various subsystems in the feedback loop. In general, the individual noise contributions are either directly measured or inferred, using different methods. In particular, these methods are:
\begin{itemize}
\item{The time-dependent part of the sensing function is measured by injecting digital signals of known shape, prior to the actuation.} 
\item{The calibration of the actuation function usually yields the largest source of errors.Until the fifth LIGO science run, the main method to measure the actuation function was the so-called free-swinging Michelson technique. Recently, a new method, called photon calibrator (PCal) has been introduced; it  uses a laser to push the end mirrors with a known radiation pressure.}
\item{The digital filters $D(f)$ are very well known functions to which we do not assign errors.}
\end{itemize}

For a full treatment of different gravitational wave interferometer calibration techniques, and the errors related to them, see \cite{LigoS5,VirgoV1,VirgoV2,Pcal,Siemens2004,Sigg2003}. Note that the time dependent part of $R(t,f)$ is slowly varying, with time scales on the order of days, while the typical signals of our interest occur on time scales of several minutes\footnote{There are other kind of longer signals, which are scientifically interesting (e.g. stochastic background, pulsars signals) but they are not considered in this work.}. By preallocating the errors due to the time dependence of the length response function, we will commit to a slight abuse of notation and write $R(t,f)=R(f)$, and include the time dependent measurement errors associated with $R(t,f)$ to the measurement of $\alpha(t)$. 
}

The transfer function $R(f)$ is a complex function. Hence, we can write it in polar form:

\beq\label{Rdefin}
R(f) \equiv A(f) e^{i \phi(f)}.
\eeq

Once the transfer function is known, the DARM can be calculated directly using Eq.~(\ref{ErrorToDarm}) from which the strain follows immediately: 

\beq\label{Strain}
 d(f) = \frac{\Delta L_{ext} }{L}
\eeq
where $L$ is the arm length of the IFO in the absence of external solicitations. 

\section{Calibration errors}\label{Sec.CalibrationErrors}

The calibration procedures are not free from systematic effects. In general the transfer function will not be known with arbitrary precision, but it will be different from the ``exact" one. These differences will be present both in amplitude and in phase:

\beq\label{Rmeasured}
R_m(f) \equiv [A +\delta A] e^{i (\phi + \delta \phi)} = \left[1 +\frac{\delta A}{A}\right]  e^{i \delta \phi(f)} R_e(f)
\eeq

Henceforth we will use an index $e$ to denote the exact length function, and all the quantities that are built from it, and an index $m$ to denote quantities which are measured, and hence affected by calibration errors (CEs).
The errors are usually reported as relative errors for the amplitude $\delta A/A$ and as the absolute ones for the phase (in radians or degrees). 

In the scenario where calibration errors are present and not negligible, the experimenter will be using the measured transfer function $R_m(f)$ and not the correct one, therefore the inferred values for the DARM and data stream will also be different from their true value.
From Eqs.~(\ref{ErrorToDarm}), (\ref{Strain}) and (\ref{Rmeasured}):

\beq\label{MeasuredData}
d_m(f) =R_m(f) \frac{e(f)}{L} = K(f) d_e(f)
\eeq
where, in order not to burden the formulae, we have introduced a function $K(f)$ that conveys the errors for both phase and amplitude:

\beq\label{Kappa}
K(f)\equiv [1 +\frac{\delta A(f)}{A(f)}]  e^{i \delta \phi(f)} 
\eeq

When a GW signal $s(f)$ and noise $n(f)$ are present in the data, they will be affected by the errors in the same way:

\beqa
d_m(f) &\equiv & n_m(f) + s_m(f) =\nonumber \\
=K(f) d_e(f)& =& K(f) \left[n_e(f) + s_e(f)\right]
\eeqa
which straightforwardly gives:

\beqa
s_m(f) &=& K(f) s_e(f) \label{MeasuredSignal}\\
n_m(f) &=& K(f) n_e(f). \label{MeasuredNoise}
\eeqa

Note that the errors do not affect what is really happening in the IFO, which is the error signal, but only the way in which this quantity is interpreted by the observers in terms of data stream.

The effects of CEs on detection statistics, and SNR, have been already the object of the work of several groups. It is known that CEs do not affect the optimal SNR \cite{Schutz2009}. This is easily verified starting from the definition of the optimal SNR $\rho$:

\beq\label{OptimalSNR}
\rho^2 \equiv 4 \int{df \frac{s(f) s^*(f)}{S(f)}}
\eeq
where we have introduced the \emph{one-sided noise spectral density} (PSD) $S(f)$, which is the Fourier transform of the noise autocorrelation function. There are several equivalent definitions for this quantity. The one we find the most useful is (see \cite{Maggiore}):

\beq
\delta(f-f') S(f) \equiv 2 \left\langle n(f) n^*(f')\right\rangle
\eeq
where the $\left\langle \,\, \right\rangle$ indicates an average over an ensemble of noise realizations.
We can easily infer the effect of CE on the noise PSD, using Eq.~(\ref{MeasuredNoise}):

\beq\label{MeasuredPSD}
S_m(f) \propto \left\langle n(f) n^*(f')\right\rangle = \left[1 +\frac{\delta A(f)}{A(f)}\right]^2 S_e(f)
\eeq
which shows how only amplitude errors affect the noise PSD.
From Eq.~(\ref{OptimalSNR}) and (\ref{MeasuredPSD}) the invariance of the optimal SNR follows nearly immediately:

\beqa
\rho_m^2 &\equiv& 4 \bigg[\int_{f_{low}}^{f_{up}}{\mbox{df} \frac{s_m(f) {s_m(f)}^*}{S_m(f) }} \bigg]=\nonumber \\
&=&  4  \bigg[\int_{f_{low}}^{f_{up}}{\mbox{df} \frac{s_e(f) {s_e(f)}^*\, [1 +\delta A/A]^2}{S_e(f)\,  [1 +\delta A/A]^2}} \bigg] =\nonumber \\
&=&4 \bigg[\int_{f_{low}}^{f_{up}}{\mbox{df} \frac{s_e(f) s_e(f)^*}{S_e(f)}} \bigg]\nonumber\\
&=&\rho_e ^2. \label{OptimalSNRUnchanged}
\eeqa
 
On the other hand, CEs do affect the actual SNR recovered by detection pipelines. In Ref.~\cite{Allen} it was theoretically calculated that the effect of CEs on the recovered SNR are of second order, for small errors. This fact was then verified experimentally, using hardware injections, during the first science run of the LIGO instruments (\cite{Brown2003}), finding that the recovered SNR depended quadratically on the time dependent part of the sensing function, $\alpha(t)$.

Theoretical approaches to the effects of CE on signal detection and template bank searches have been pursued in Refs.~\cite{Lindblom2008, LindblomAug2009,LindblomSept2009}. In \cite{Bose2005} these studies were extended to include the effects of parameter estimation for various kind of signals. A theoretical study that makes use of Bayesian analysis is being performed by one of the authors \cite{Vitale2011prep}.

Without going into details, it seems clear that calibration errors have the potential to impact the measurement of all of the source parameters -- masses, sky location, distance, inclination and orientation -- because of the complicated correlations that exist between these parameters.  Therefore, precisely evaluating the impact of calibration errors requires a careful numerical analysis that coherently fits all parameters simultaneously, and this is the analysis we present in subsequent sections.  

Here we rely on approximations to crudely estimate the most significant biases due to possible calibration errors.  The intrinsic parameters (the two component masses, and potentially spins, though we do not consider these here) leave a very strong signature on the phase evolution of the gravitational waveform, and are primarily measured through phase rather than amplitude information.  The sky location can be estimated by timing triangulation between the arrival times of the GW signal at different detectors.  The inclination and orientation angles are functions of the relative signal amplitudes and phase shifts at the detectors, while the distance is given by the overall signal amplitude once other parameters are known.  These angles and distance are strongly correlated with each other, but relatively weakly correlated with the intrinsic parameters.

Calibration errors can be divided into three types: timing errors, amplitude errors and frequency-dependent phase errors, and one can estimate the permissible ranges on the three error types subject to the condition that systematic biases must remain below statistical measurement uncertainties. 

\begin{enumerate}
\item{Timing errors.}  These primarily affect sky localization by influencing timing triangulation, and can be seen as a special case of phase errors described below (phase errors with linear dependence on the frequency).
A source can be timed to a $O(1/\mbox{SNR})$ fraction of a wave cycle, with the best timing happening at the ``bucket'' of the noise spectrum, around 100 Hz.  Thus, we may expect timing accuracies of order a millisecond.  Meanwhile, the typical baseline (separation between detectors) is of order 10 milliseconds of light travel time, leading to statistical measurement uncertainties of order 10 degrees for a pair of detectors.  Timing errors will, therefore, become significant relative to measurement uncertainties only if they constitute a significant fraction of a millisecond, and calibration-induced biases should be negligible for timing errors of less than $\sim 0.1$ ms.  (Note, however, that measurement errors improve with more detectors, so an expansion of the detector network will increase constraints on timing errors.) In this work we will not consider this kind of errors, as the actual timing errors measured by the calibration teams~\cite{LigoS5,VirgoV1,VirgoV2} are much smaller than the values which might lead to large biases.

\item{Amplitude errors.}  If constant amplitude errors lead to a fixed scaling of the measured amplitude in all detectors, they would only affect the distance estimate and none of the other parameters.  Distances are not particularly well-measured by GW networks, with typical fractional uncertainties of perhaps $300/\mbox{SNR}$\%, so for an individual source, amplitude calibration errors of under 20\% should not lead to dominant systematic errors, except for the loudest events.\footnote{It is worth pointing out, however, that if amplitude calibration errors stay constant over the run, these distance biases would be constant unlike the randomly fluctuating measurement uncertainties, so they could have a pernicious effect on analyses that combine observations of multiple sources to study cosmology \cite{DelPozzo,TaylorEtAl:2011}.}  Of course, amplitude calibration errors will not be identical in the various detectors, so inclination and orientation will be affected along with distance, but due to the difficulty of measuring these parameters precisely, similar constraints apply.  Frequency-dependent amplitude errors should not significantly influence parameter estimation for nonspinning signals, since estimates will primarily be sensitive to a (noise-weighted) average amplitude; however, spin measurements are sensitive to modulations of signal amplitude which could mimic the effects of orbital precession, hence such errors could cause more problems if spin parameters are also being estimated. 

\item{Frequency-dependent phase errors.}  Frequency-dependent phase errors are, perhaps, the most dangerous of all, since they can influence the measurements of the binary's intrinsic parameters.  Such errors can mimic the effects of different post-Newtonian corrections to the phase evolution, leading to systematic biases in the measurements of the masses.  However, these phase errors are localized in frequency and do not accumulate over the inspiral.  Therefore, sensitivity to these errors is limited by the overall measurement uncertainty on the waveform phase, which is expected to be on the order of $1/\mbox{SNR}$ of a cycle at the bucket, and worse elsewhere.  Therefore, frequency-dependent phase errors of less than $\sim 10$--$20$ degrees should not lead to significant biases for all but the strongest signals.
\end{enumerate}

The rest of the paper is dedicated to the systematic study in the context of Bayesian inference of the combined effects of phase and amplitude calibration errors on parameter estimation for GW signals emitted during the in-spiral of compact binary systems whose components are not spinning. 

\section{Bayesian model selection and parameter estimation}\label{Sec.BayesianPE}

An excellent introduction to Bayesian model selection, and its application to GW detection an parameter estimation can be found in \cite{Veitch2010}. In this paragraph we will only summarize the main results and nomenclature we will use in the remainder of this work. 

Given a set of data $\vec{d}$ and some prior information $I$, the probability for a model (or hypothesis) $\H_i$ is given by Bayes' theorem:

\beq\label{PosteriorH}
P(\H_i|{\vec d},I)=\frac{P(\H_i|I)P({\vec d}|\H_i,I)}{P({\vec d}|I)},
\eeq
where $P(\H_i|I)$ is the prior probability for the hypothesis $\H_i$, and $P({\vec d}|\H_i,I)$ is the posterior probability for the data given that the hypothesis $\H_i$ is true, also called the \emph{likelihood} for the data.
The factor in the denominator, $P({\vec d}|I)$, is the marginal probability for the data, integrated over the different hypotheses or models.

Without enumerating all the different models, we can calculate the relative weight between two of them (the \emph{odds ratio}), using  Eq.~(\ref{PosteriorH}). More precisely, the odds ratio of a model $\H_i$ and a model $\H_j$ is:

\beq\label{Odds}
O_{i,j}=\frac{P(\H_i|I)}{P(\H_j|I)}\frac{P({\vec d}|\H_i,I)}{P({\vec d}|\H_j,I)}=\frac{P(\H_i|I)}{P(\H_j|I)}B_{ij}\,,
\eeq
where we have introduced the \emph{Bayes factor} $B_{ij}$, or ratio of likelihoods, between model $\H_i$ and model $\H_j$.
Note that the marginal probability for the data, $P({\vec d}|I)$, cancels out when the ratio is calculated.

In a typical scenario, the GW signal will depend on a set of unknown parameters $\vec{\theta}$ that we want to estimate. These can be both extrinsic parameters, such as the position of the GW source on the sky, and intrinsic parameters, such as the mass of the component stars. If we indicate with $\Theta$ the parameter space in which $\vec{\theta}$ dwells, we can obtain the likelihood for the data given the generic model $\H$ by marginalization of the likelihood given a particular realization of $\vec{\theta}$, and obtaining the \emph{evidence} $Z_\mathcal{H}$:

\beq\label{Evidence}
Z_\H= P({\vec d}|\H,I) =\int_\Theta p(\pvec|\H,I)p({\vec d}|\H,\pvec,I)d\pvec,
\eeq
where we have introduced the prior probability distribution $p(\pvec| \H,I)$ for the parameters $\pvec$ over the parameter space.
From the evidence, the posterior distributions for the parameters $\pvec$ given the data are easily obtained using Bayes' theorem:
\beq
p(\pvec|{\vec d},\H,I) = \frac{p(\pvec|\H,I) p({\vec d}|,\pvec,\H,I)}{Z_\H}
\eeq

Given the high dimensionality and the analytical form of the functions involved, the integral (\ref{Evidence}) cannot be calculated analytically, and one has to rely on numerical methods. For our computations, we relied on the Nested Sampling algorithm (\cite{Skilling}) in the form in which it has been implemented for the LIGO Algorithm Library (LAL) \cite{LAL} by Veitch and Vecchio \cite{Veitch2010}.

In what follows, we will consider two hypotheses: (i) $\H_N$ will be the hypothesis according to which the data consist solely of noise; (ii) $\H_S$ will be the hypothesis that the data consist of noise plus a GW:

\beqa
\H_N &\rightarrow  & d(f) \equiv n(f) \\
\H_S &\rightarrow & d(f) \equiv n(f) + s(f,\pvec)
\end{eqnarray}
where we have made explicit the signal dependence on the unknown parameter vector $\pvec$.
If we assume that the noise in the IFO is stationary and Gaussian \footnote{This is not true in general, as the noise in the IFOs is a combination of smaller Gaussian fluctuations and larger non-Gaussian outliers (``glitches'' in the data). The use of coincident requirements between different sites and a whole set of data quality and vetoes procedures help reducing the number of glitches \cite{Gonzalez,AllenChi2,AbbottAl}. New techniques are being developed to deal with residual non-Gaussianity \cite{Biswas}. For simplicity, in this work we will assume that the candidates events which survive all of these checks are buried in Gaussian noise.} the likelihood for the data for the two models can be written as:

\beqa
p(d |\pvec,\H_N,I) &\propto & e^{-\langle d(f) | d(f) \rangle/2}  \label{LikelihoodHN}\\
p(d |\pvec,\H_S,I) &\propto & e^{-\langle d(f) -h(\pvec) | d(f) -h(\pvec) \rangle/2}\,, \label{LikelihoodHS}
\eeqa
where $h(f,\pvec)$ is the GW signal, and we have defined a noise-weighted inner product:

\begin{equation*}
\langle a(f), b(f)\rangle \equiv 2 \Re \bigg[\int_{f_{low}}^{f_{up}}{\mbox{df} \frac{a(f) b(f)^* + a(f)^* b(f)}{S(f) }} \bigg]
\end{equation*}

Once the analysis is done for a given data stream, one is provided with two pieces of information:

\begin{itemize}
\item The Bayes factor between the models $\H_S$ and $\H_N$ (BSN for Bayes Signal vs Noise) which tells how confident we are that there is a signal buried into the noise.
\item The posterior distributions for the unknown parameter on which the signal (if present) depends, which allow estimates for the physical and extrinsic parameters of the GW source.
\end{itemize}

The method is easily generalized to the case where a coherent analysis is being performed, using a network of several IFOs. 
If we indicate with $d^{(J)}(f)$ the data stream in the $J$-th detector, the likelihood of having a signal or only noise in the $J$-th detector will be exactly the same as in Eqs.~(\ref{LikelihoodHN}) and (\ref{LikelihoodHS}), with $d \leftrightarrow d^{(J)}$. If the detectors are far enough apart that the noise in one is not correlated with the noises in the others, the likelihood for each IFO is statistically independent of the likelihood for the other instruments, and a joint likelihood can be built just multiplying the single IFO expressions:

\beq\label{LikelihoodTotal}
p({\vec d}|\pvec,\H_k,I)=\prod_{(J)}p({\vec d}^{(J)}|\pvec,\H_k,I)\,,
\eeq
with $k=N$ or $k=S$. Eq.~(\ref{LikelihoodTotal}) can be used to calculate the network evidence, and perform coherent analysis.

\section{Method}\label{Sec.Method}

\subsection{Analysis and Noise Model}\label{Sec.Noise}
\begin{figure}[h]
\includegraphics[scale=\SkyFigscale]{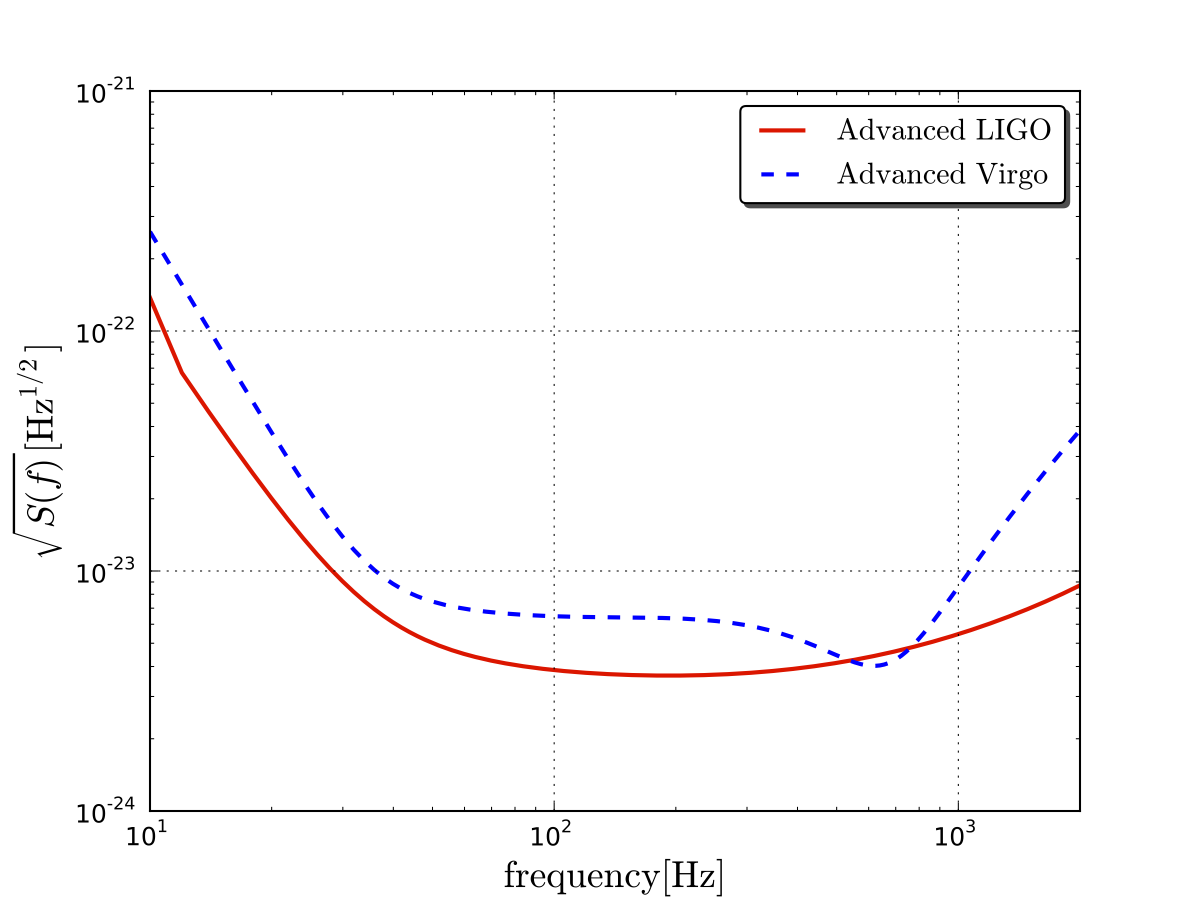}
\caption{(color online) The high-power, zero-detuning noise curve for Advanced LIGO (red continuous line), and the BNS-optimized Advanced Virgo noise curve  (blue dashed)}\label{Fig.Noises}
\end{figure}

We have tested the effects of CE on PE using software injections, i.e.~artificially adding signals of known shape into simulated noise, for a network consisting of the two advanced versions of the LIGO and Virgo instruments.
We have used the analytical expressions for the noise spectral densities as coded in LAL \cite{LAL}. The square root of $S(f)$ for advanced LIGO and Virgo is shown in Fig.~\ref{Fig.Noises}.

To be more precise, for each IFO, a GW signal $s(f)$ is added to a stream of noise generated using the designed noise PSD for that IFO, $n^{(J)}(f)$ to form the data vectors

\beq\label{DataExactJ}
d^{(J)}_e\equiv s_e^{(J)}(f) + n_e^{(J)}(f),
\eeq
that are combined to form a joint likelihood, Eq.~(\ref{LikelihoodTotal}), which is evaluated by the Bayesian pipeline. The subscript $e$ indicates that the transfer function used to create the stream is the exact one, $R_e(f)$.
The final outcomes of this analysis will be the $\mbox{BSN}_e$ (logarithmic Bayes' factor of the signal hypothesis vs the noise hypothesis) and the posterior distributions of all the component of $\vec{\theta}$, from which the mean $\bar{\theta}_e^\alpha$, standard deviation $\Delta \theta_e^\alpha$, as well as the median and higher moments of the distribution for the parameter $\theta^\alpha$ can be calculated.

Once the \emph{exact} analysis is completed, we proceeded with a similar analysis in which we artificially introduced calibration errors on signal and on noise as in Eqs.~(\ref{MeasuredSignal}) and (\ref{MeasuredNoise}). We then compared the $\mbox{BSN}_e$ and posterior distributions obtained from our pipeline in the two cases. We kept fixed all relevant parameters of the injection and of the noise generation. The only difference between the two datasets are the presence of calibration errors in one of them.

In the next few subsections we will discuss in detail which GW model waveforms have been used and how the calibration error curves have been generated.

\subsection{Waveforms and parameter space}

When software injections are used to test a parameter estimation pipeline, there are three major factors to take into account: (i) the signal being injected, (ii) the waveform used to recover the signal (known as template), and (iii) the noise added to the signal. The noise model we employed has been described in Sec.~\ref{Sec.Noise}, hence in this section we will proceed in the description of (i) and (ii).

The waveform models we used for injections belong to the Effective One Body (EOB) family \cite{Damour2001,Buonanno1999,Buonanno2000,Damour2000,Buonanno2006,Damour2007,Buonanno2007,Damour2008,Damour2008b}.

Without entering into details, which can be found in the references above, the main idea behind the EOB approach is to treat the two-body problem as an effective one-body problem, as if a mass equal to the reduced mass of the system were moving in some effective space-time metric \cite{Buonanno2000}.
The EOB's main ingredient is the effective Hamiltonian, from which the evolution of the radial and angular coordinates, as well as their momenta,  can be calculated using Lagrange equations. This allows to write the GW signal, as a function of the reduced time $\hat t \equiv t/M$ ($M$  being the total mass of the binary system) as:
\beq\label{Eq.EOB}
h(\hat t) = v_{\omega}^2(\hat t) \cos(\varphi(\hat t))
\eeq
where $ v_{\omega}$ is a power of the angular velocity, obtained deriving the phase with respect to the reduced time: $$ v_{\omega} \equiv \left(\frac{d\phi}{d\hat t}\right)^\frac{1}{3}$$ and $\varphi(\hat t)$ is twice the orbital phase: $\varphi \equiv 2 \phi$.

It is important to note that using a template family which is different from the injected signal's may introduce a bias in the recovered posteriors for the parameters \cite{Lindblom2008}. However, let us remember that in this work we are not interested in the \emph{absolute} performance of the code, or in the match between the injected and recovered parameters. What we want to measure, instead, are the effects of CEs, i.e. how much the posteriors are affected by the presence of CEs. 
Now, as we are dealing with small errors, it makes sense to assume that even if a bias was introduced, it would be the very similar while recovering $s_e(f)$ or $s_m(f)$, and will become negligible when the difference \mbox{$\theta^\alpha_m(f) - \theta^\alpha_e(f)$} is taken, which we use to quantify the shift introduced by the CE. 
With this in mind, we have chosen to use a frequency domain template, the Taylor F2 discussed here below, because it is known analytically, and no differential equations have to be solved, thus the performance of the code is greatly improved compared to more sophisticated models.

The TaylorF2 waveform \cite{Buonanno2009} is calculated starting from the time-domain Post-Newtonian (PN) approximation of the signal:
\beq\label{Eq.TT2}
h(t)=  v^2(t) \cos(\varphi(t)) 
\eeq
which looks equal to Eq. (\ref{Eq.EOB}). The difference is that now the amplitude and the (double of the) orbital phase are calculated starting from PN expansions of the energy flux and luminosity, and assuming that the adiabatic approximation holds; and are known functions of the system's parameters (see \cite{BlanchetLiving} and references therein). The Fourier transform of Eq. (\ref{Eq.TT2}) can be analytically calculated using the so-called \emph{stationary phase approximation} \cite{Damour2005}, which consist in developing the phase of the signal around its stationary point. The final result is:
\beq\label{Eq.TF2}
 h(f)= \frac{Q(\theta,\phi) M^\frac{5}{6}}{\pi^\frac{2}{3} D} \sqrt{\frac{5\eta}{24}}\, f^{-\frac{7}{6}} \,e^{i\,\psi(f)}
\eeq
where the phase is given at the 3.5 PN order by:
\begin{eqnarray}
 \psi(f)&=& 2\pi f t_0 + \phi_0 - \frac{\pi}{4} + \frac{3}{128 \eta v^5}  \sum_{k=0}^7 \alpha_k v^k
\end{eqnarray}
and $v\equiv (\pi M f)^{\frac{1}{3}}$. The coefficients $\alpha_i$, that depend on the total and symmetrized mass, can be found in \cite{Arun2005,ArunCorrection}.
The function $Q(\theta,\phi)$ depends on the coordinates of the source in the detector frame. When more IFOs are used to perform coherent analysis, one has to use a common frame, and the functions $Q$ will depends both on the spherical coordinates of the source in the common frame and on the Euler angles that rotate the detector frame to the common frame \cite{Vitale2011}.

The signal emitted by a binary system with zero eccentricity\footnote{By the time the system's frequency enters the Ligo-Virgo bandwidth, most of the eccentricity will have been radiated away~\cite{Peters1964}, which is why it is usually neglected in the LIGO-Virgo literature.} and nonspinning components will depend on nine parameters:

\begin{itemize}
\item A reference time (usually the detection time, or the coalescence time) and the phase the waveform had at that time: $t_0$ and $\phi_0$.
\item  The total mass $M=m_1 +m_2$, and the symmetric mass ratio $\eta =\frac{m_1 m_2}{(m_1 +m_2)^2}$. The \emph{chirp mass} $\mathcal{M} = \eta^\frac{3}{5} M$ is often used instead of the total mass, as it is generally the best-determined variable.
\item The luminosity distance of the system, D.
\item  The polarization angle, $\psi$ \cite{Thorne300}.
\item The angle formed between the line of sight and the system orbital angular momentum, $\iota$.
\item The coordinates of the sources in the common frame, right ascension (RA) and declination (dec).
\end{itemize}

The injections were collected in three catalogs, each one representative of a different kind of binary system, composed of two neutron stars (BNS), two black holes (BBH) or a neutron star and a black hole (BHNS). We will denote those catalog as $\mathcal{E}_j \mbox{ with j=BNS, BBH, BHNS}$.
We have assumed that a NS has a mass in the range $[1.4, 2.3] M_\odot$ and BH in the range $[9.0, 11.0] M_\odot$. While there are scientific reasons to believe that the mass of a NS is in that range \cite{BNSMasses}, for the BH the range of allowed masses is much broader, going from a few solar masses up to thousands of solar masses for the black holes in the center of the galaxies. We have chosen a range centered around $10 M_\odot$ as that is the value most often used in the GWs data analysis literature.
For each catalog, the distances of the signals were randomly drawn from ranges chosen in such a way that the corresponding SNR would have values like those we expect from detections with the Advanced Interferometers. 
The corresponding mass for the two objects, and the distance, for binary systems in the classes above are given in Table~\ref{Table.Ranges}.

\begin{table}[htb]
\begin{tabular}{|c|c|c|c||}
\hline
 & $m_1$ & $m_2$ &$D$   \\ 
 \hline
BNS & [1.4, 2.3] $M_\odot$ & [1.4, 2.3] $M_\odot$ &  [150, 220] Mpc\\
 BBH & [9.0, 11.0] $M_\odot$ & [9.0, 11.0] $M_\odot$ & [700, 1000] Mpc \\
 BHNS &  [1.4, 2.3]$ M_\odot $& [9.0, 11.0] $M_\odot$ &  [300, 500] Mpc\\
 \hline
\end{tabular}
\caption{Mass and distance ranges for the systems considered }\label{Table.Ranges}
\end{table}

Each catalog was filled with 250 signals, whose corresponding masses and distances were generated by sampling uniform distributions on the intervals indicated in Table~\ref{Table.Ranges}.
The other parameters, the sky positions of the sources as well as the polarization and inclination angles, were generated by sampling uniform distributions on the 2-sphere.

It is worth noticing that the only things that change while going from the I-th event of one catalog to the I-th event of another catalog are the masses and distance, while the other parameters are the same. This implies that we can use this work to quantify the effects of CEs on signals having comparable masses but different positions, polarization, inclination and distances (this is done analyzing each catalog) and the effects on signals having the same positions, polarizations and inclination, but different masses (this is done comparing a catalog with the others).

\subsection{Generating calibration errors}

It is a reasonable assumption that, at the beginning of the advanced detectors era, the errors in the calibration process will not be much different from what they were during the last part of the initial detectors era \cite{GonzalezPrivate,RollandPrivate}. 

In order to have a good statistical sample, and take into account possible slow time variation, due to $\alpha(t)$, we have generated 10 different error curves for each IFO, for both phase and amplitude. 

Each of these curves was created using the following method:

\begin{itemize}
\item Read the typical width of the 1-sigma calibration errors curves during the last stages of the Initial detector era
\item Draw ~15 points in the frequency space, uniformly in $\log f$, from Gaussian distributions with zero (one) mean for the phase (amplitude) uncertainties
\item Fit these points with a polynomial of degree 7 to obtain a smooth parametrized curve.
\end{itemize}

\begin{figure}[htb]
\includegraphics[scale=\Figscale]{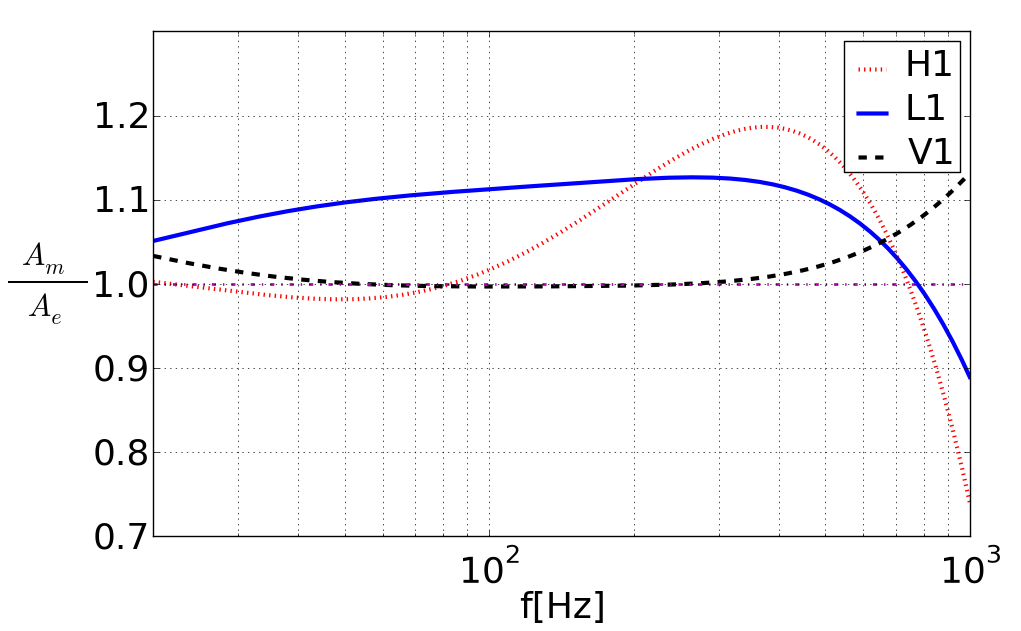}\\
\includegraphics[scale=\Figscale]{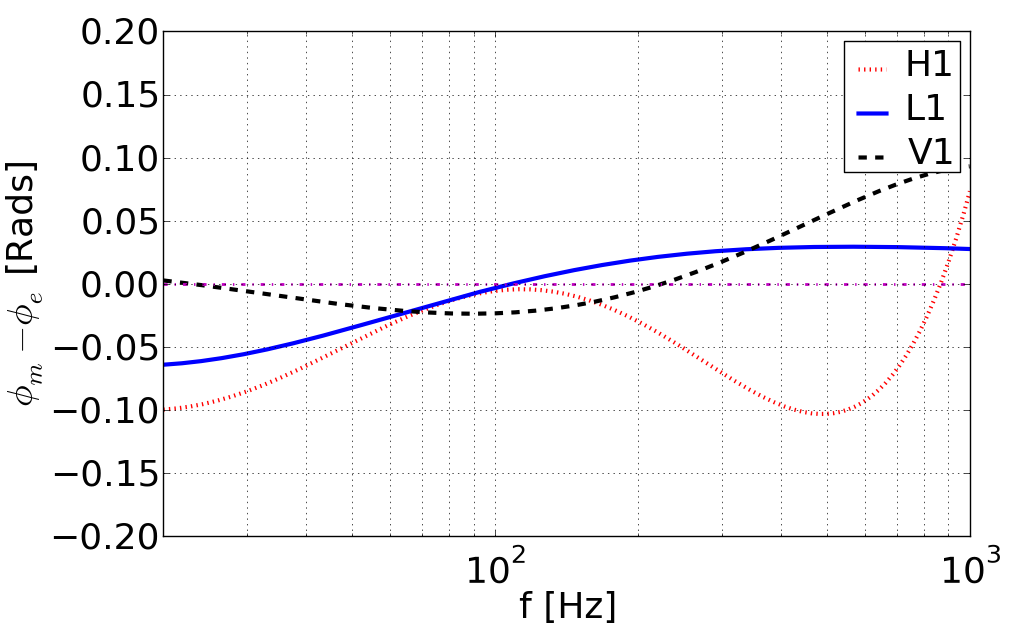}
\caption{(color online) The first CE realization for the amplitude (top) and phase (bottom).}\label{Fig.Errors1}
\end{figure}

The aforementioned process was repeated using different seeds for the initialization of the random number generator so to obtain different curves. An instance of the different realizations we generated is shown in Fig.~\ref{Fig.Errors1}. The interested reader is referred to the Appendix, Figs.~\ref{Fig.Errors2} to \ref{Fig.Errors10} for an overview of all the realizations.
The values of the widths we have used are given in the Table~\ref{Table.Widths}, and refer to the values estimated during the S5 science run for LIGO and the third science run for VIRGO \cite{LigoS5,VirgoCali}. Adopting the LIGO-Virgo conventions, we will use the label {\bf L1} for the LIGO instrument in Livingston,  {\bf H1} for the LIGO detector in Hanford and {\bf V1} for Virgo.

\bwt
\begin{table}[h!b!p!]
\begin{tabular}{|c|c|c|c||c|c|c|c|}
\hline
&\multicolumn{3}{c||}{Amplitude errors (\%)} & \multicolumn{4}{c|}{ Phase errors (Deg)} \\
 & 40-2000Hz & 2-4KHz & 4-6KHz  & \multicolumn{2}{c|}{40-2000Hz} & 2-4KHz & 4-6KHz \\ 
 \hline
 H1 & 10.4 & 15.4 & 24.2 & \multicolumn{2}{c|}{4.5} & 4.9 & 5.8 \\
 L1 & 10.1 & 11.2 & 16.3 & \multicolumn{2}{c|}{3.0} & 1.8 & 2.0 \\
 \hline \hline
 & 40-2000Hz & 2-4KHz & 4-6KHz  & 40-500Hz& 500-2000 Hz & 1-2.8KHz & 2.8-6KHz \\ 
 \hline
 V1 & 10.0 & 10.0 & 20.0 & 2.29+2.87$\cdot 10^{-3}$ f &0.5729 +6.3$\cdot 10^{-3}$ f & 6.87 & 2.53$\cdot 10^{-3}$ f \\
 \hline
\end{tabular}
\caption{The widths used for the error curves generation. The phase error width for Virgo depend on the frequency f \cite{VirgoCali}}\label{Table.Widths}
\end{table}
\ewt

We will indicate with $\delta A_i/A_i$ and $\delta \phi_i$ the $i$-th realization of the amplitude and phase errors. 
Note that drawing the points uniformly in $\log f$ is equivalent to assuming that there is a correlation length between the errors at different frequencies which increases linearly with the frequency. We will consider different possibilities in future work, even though the consistency of the results we have obtained using the various curves in this work (see Sec.~\ref{Sec.Compare} below) suggests the results are not extremely dependent on the exact shape of the calibration error curve.

\section{Results}\label{Sec.Results}

\subsection{Effects on Parameter Estimation}

\begin{figure}[htb]
\includegraphics[scale=\Figscale]{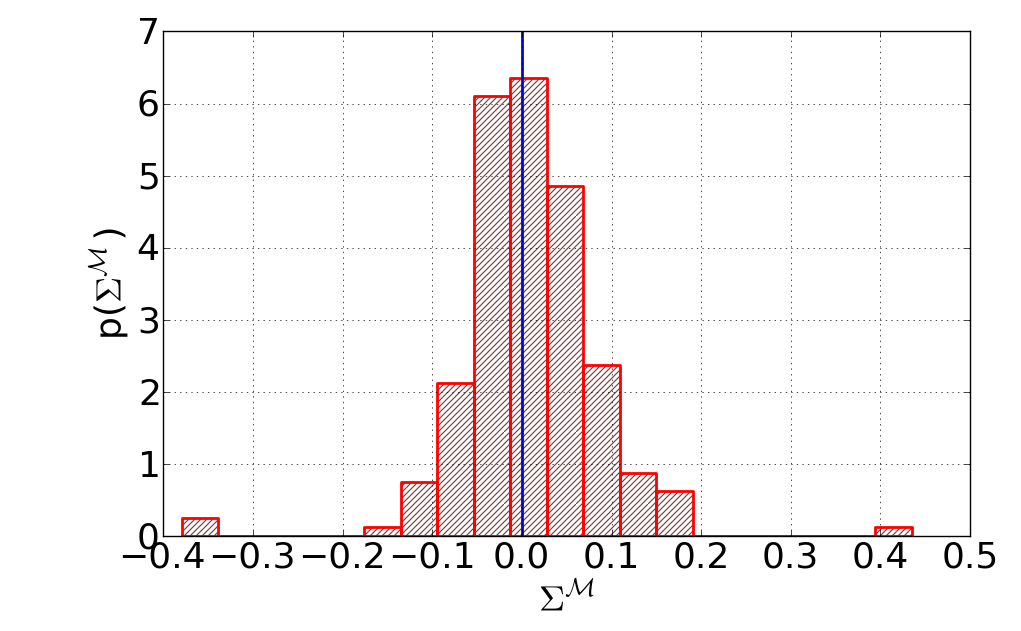}
\caption{(color online) The distribution of $\Sigma^{\mathcal{M}}$ for the signal in the BHNS catalog, using the first CE realization. The vertical blue line correspond to a null shift.}\label{Fig.EffMBHNS}
\end{figure}

Because of the different ranges in which each parameter can vary, we have normalized the difference in the means or medians of the parameters inferred from runs with and without calibration errors by their  standard deviation.  More precisely, if $\underline{\theta}_e^\alpha$ and $\Delta \theta_e^\alpha$ are the median and standard deviation of the parameter $\theta^\alpha$ we would measure for a given signal if we knew the exact transfer function, while  $\underline{\theta}_m^\alpha$ is the median we measure when CEs are present, we can build the quantity:
\beq\label{EffectSizeGeneric}
\Sigma^\alpha \equiv \frac{ \underline{\theta}_m^\alpha - \underline{\theta}_e^\alpha}{\Delta \theta_e^\alpha},
\eeq
the meaning of which is clear: it measures the shift introduced in the estimate of $\theta^\alpha$ by the CEs in units of standard deviations calculated from the probability distribution for the same parameter in the absence of CEs.
For each injection, say the $i$-th, in the catalog $\mathcal{E}_j$ we can calculate the quantity (\ref{EffectSizeGeneric}):
\beq\label{EffectSizeIJ}
\Sigma_{i}^\alpha \equiv \frac{ \underline{\theta_i}_m^\alpha - \underline{\theta_i}_e^\alpha}{\Delta {\theta_i}_e^\alpha}\,,\mbox{  i=1..250}
\eeq
where $\underline{\theta_i}$ is the median for parameter $\theta_i$. We also compute distributions for this quantity for all of the injections in the catalog, and for all the parameters of the model waveform. The resulting distributions will look in general similar to Fig.~\ref{Fig.EffMBHNS} which shows the histogram for the chirp mass $\mathcal{M}$ measured using the BHNS catalog\footnote{The results are similar for the three catalogs. To avoid having too many figures, we have chosen to show plots only for the BHNS catalog. It is understood that one would get very similar plots for the other two catalogs. } and the first CE realization.

\begin{figure}[htb]
\includegraphics[scale=\Figscale]{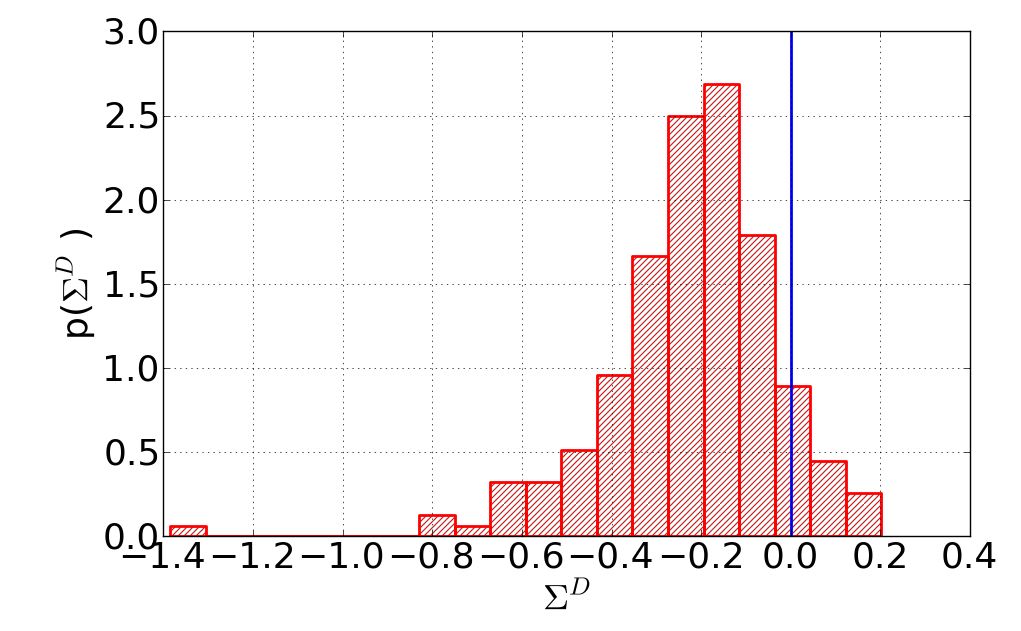}
\caption{The distribution of $\Sigma^{D}$ for the signals in the BHNS catalog, using the first CE realization. The vertical blue line correspond to a null shift.}\label{Fig.EffDBHNS}
\end{figure}

Note that the distribution for $\Sigma^{\mathcal{M}}$ looks quite symmetric and well centered around zero, meaning that there is not a net bias introduced by CEs but, instead, some of the injections in the catalog acquire a positive bias while others a negative one. We found that this behavior is common to all parameters except for the distance. The reason is easy to understand: with other parameters fixed, the distance is inversely proportional to the amplitude of the signal, and is therefore directly affected by the amplitude errors of the transfer function. As an example, in the same CEs realization, Fig. \ref{Fig.Errors1}, the amplitude errors are positive for the three IFOs. The over-estimated amplitudes result in an under-estimate of the distance, so the source is inferred to be closer than in the absence of CEs, Fig.~\ref{Fig.EffDBHNS}.

As a summary for our results, we will report the mean $\overline{\Sigma}$, and standard deviation $\Delta \Sigma$, of the distribution for $\Sigma^\alpha$, together with the median,  $\underline{\Sigma} $ , the $5^{\mbox{th}}$ and $95^{\mbox{th}}$ percentiles, for each parameter and each catalog, averaged over the 10 CE realizations.
It is important to remember that $\Sigma$s represent the effect of systematic errors and are not normally distributed. In particular $2 \Delta\Sigma$ does not to contain $\sim66\%$ of the results. 
The results are summarized in Tables~\ref{Table.SummaryBNS},~\ref{Table.SummaryBBH} and~\ref{Table.SummaryBHNS}. 

\begin{figure}[htb]
\includegraphics[scale=0.35]{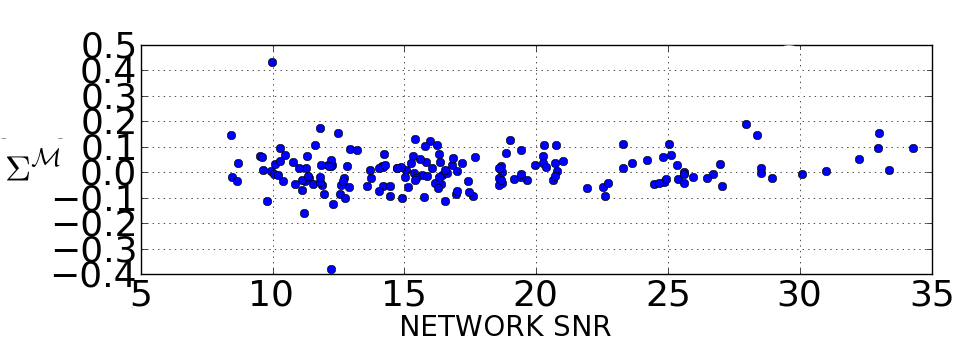}
\caption{(color online) $\Sigma^{\mathcal{M}}$ for the signal in the BHNS catalog, using the first CE realization, plotted against the SNR. The fact that the spread does not increase with the SNR implies that the net bias $\underline{\theta}_m^\mathcal{M} - \underline{\theta}_e^\mathcal{M}$ decreases with the SNR.}\label{Fig.effect_vs_snr}
\end{figure}

The distribution for $\Sigma^\alpha$ has been calculated using only the injections whose network SNR is greater than 8, which we used as a proxy for the sensitivity of GW searches.
It is important to note, however, that excluding those injection (which are $\approx 20\%$ of the total number) does not affect our analysis in a significant way. On the contrary, those weak signals would produce posterior distributions with large standard deviations, and thus small $\Sigma$s, reducing the spreads of the $\Sigma$s.
It is interesting to check whether the net bias (not weighted by the standard deviation) is a function of the SNR. 
At first one might think that calibration-induced systematic errors must be not dependent on the SNR, as this is the case, for example, for the bias introduced by using wrong templates \cite{Cutler}. When it comes to calibration errors, however, there is an important difference: not only the template but also the noise is affected (see Eq. \ref{MeasuredNoise}). However, in Fig. \ref{Fig.effect_vs_snr}, we show  $\Sigma^{\mathcal{M}}$ for the same signal as in Fig. \ref{Fig.EffMBHNS} plotted against SNR. As the random errors decrease with the SNR, the fact that the ratio between the bias and the standard deviation (i.e.,  $\Sigma^{\mathcal{M}}$) is not increasing with the SNR implies that the net bias is also decreasing with the SNR. We conjecture that this is due to an important difference between systematic errors induced by theoretical waveform differences and calibration errors: in the latter case, not only the template but also the noise is affected (see Eq. (3.6) and Eq. (11) of \cite{Cutler}). Finally, we point out that our procedure for estimating bias as the difference in the medians between posterior samples in the error-free and CE-affected runs includes two effects: a genuine systematic bias and a Monte Carlo sampling fluctuation due to finite sample statistics. The latter will scale as $ \mbox{SNR}^{-1}$, and could dominate the estimated bias when the bias induced by calibration errors is very small. Thus, our quoted biases represent a conservative upper limit on CE-induced systematic errors. We will study these issues in a follow-up project.

\begin{table}[htb]
\begin{tabular}{|c||c|c|c|c|c|}
\hline
 & $\overline{\Sigma} $& $\Delta \Sigma$ & $\underline{\Sigma} $ & $5^{th}$  &$95^{th}$  \\ 
 \hline
 $\mathcal{M}$ & -7.29$\cdot 10^{-3}$& 2.07$\cdot 10^{-1}$ &-2.35$\cdot 10^{-3}$ &-2.42$\cdot 10^{-1}$ &2.02$\cdot 10^{-1}$\\
  \hline
$\eta$ & -1.62$\cdot 10^{-2}$ &1.92$\cdot 10^{-1}$ &3.20$\cdot 10^{-3}$ &-2.49$\cdot 10^{-1}$& 1.68$\cdot 10^{-1}$\\
 \hline
 RA & 1.21$\cdot 10^{-2}$ &4.96$\cdot 10^{-1}$ &-1.64$\cdot 10^{-3}$ &-3.79$\cdot 10^{-1}$& 4.41$\cdot 10^{-1}$\\
 \hline
 dec. & 1.56$\cdot 10^{-2}$& 4.48$\cdot 10^{-1}$& -1.20$\cdot 10^{-2}$ &-4.61$\cdot 10^{-1}$& 5.11$\cdot 10^{-1}$\\
 \hline
$\psi$ & -7.57$\cdot 10^{-4}$& 3.51$\cdot 10^{-2}$& -6.80$\cdot 10^{-4}$ &-5.40$\cdot 10^{-2}$ &5.48$\cdot 10^{-2}$\\
 \hline
 $\phi_0$&-2.43$\cdot 10^{-3}$& 3.35$\cdot 10^{-2}$ &-1.28$\cdot 10^{-3}$ &-5.62$\cdot 10^{-2}$ &4.78$\cdot 10^{-2}$\\
 \hline
 $t_0$ &-1.93$\cdot 10^{-3}$& 4.47$\cdot 10^{-1}$ &-3.20$\cdot 10^{-4}$ &-3.87$\cdot 10^{-1}$& 3.91$\cdot 10^{-1}$\\
 \hline
 $D$ & -1.37$\cdot 10^{-2}$& 2.33$\cdot 10^{-1}$& -4.94$\cdot 10^{-3}$ &-2.89$\cdot 10^{-1}$& 2.29$\cdot 10^{-1}$\\
 \hline
 $\iota$ &-1.52$\cdot 10^{-2}$& 4.58$\cdot 10^{-1}$& -9.08$\cdot 10^{-4}$ &-5.98$\cdot 10^{-1}$& 5.35$\cdot 10^{-1}$ \\
 \hline
\end{tabular}
\caption{The mean $\overline{\Sigma}$, standard deviation $\Delta \Sigma$, median $\underline{\Sigma} $, 50th and 95th percentile of $\Sigma$ for all the parameters using the BNS catalog. These numbers are obtained by averaging over ten CEs realizations. All the quantities are pure numbers (remember the definition Eq.~\ref{EffectSizeIJ} of $\Sigma$).}\label{Table.SummaryBNS}
\end{table}

\begin{table}[htb]
\begin{tabular}{|c||c|c|c|c|c|}
\hline
 & $\overline{\Sigma} $& $\Delta \Sigma$ & $\underline{\Sigma} $ & $5^{th}$  &$95^{th}$  \\ 
 \hline
 $\mathcal{M}$ & 1.72$\cdot 10^{-3}$& 8.48$\cdot 10^{-2}$& 3.06$\cdot 10^{-3}$& -1.18$\cdot 10^{-1}$& 1.20$\cdot 10^{-1}$\\
  \hline
$\eta$ &-1.82$\cdot 10^{-4}$& 1.06$\cdot 10^{-1}$& 2.09$\cdot 10^{-3}$& -1.21$\cdot 10^{-1}$& 1.20$\cdot 10^{-1}$\\
 \hline
 RA &6.04$\cdot 10^{-3}$& 3.43$\cdot 10^{-1}$& 1.58$\cdot 10^{-3}$& -4.00$\cdot 10^{-1}$& 4.06$\cdot 10^{-1}$\\
 \hline
 dec. & -3.72$\cdot 10^{-2}$& 3.96$\cdot 10^{-1}$& -2.19$\cdot 10^{-2}$& -4.89$\cdot 10^{-1}$& 3.95$\cdot 10^{-1}$\\
 \hline
$\psi$ & 4.52$\cdot 10^{-5}$& 4.16$\cdot 10^{-2}$& 8.63$\cdot 10^{-4}$& -5.31$\cdot 10^{-2}$& 5.12$\cdot 10^{-2}$\\
 \hline
 $\phi_0$ & -3.67$\cdot 10^{-4}$& 4.18$\cdot 10^{-2}$& -2.26$\cdot 10^{-4}$& -4.96$\cdot 10^{-2}$& 4.99$\cdot 10^{-2}$\\
 \hline
 $t_0$ &-3.14$\cdot 10^{-2}$& 3.76$\cdot 10^{-1}$& -7.10$\cdot 10^{-3}$& -3.03$\cdot 10^{-1}$& 2.48$\cdot 10^{-1}$\\
 \hline
 $D$ & -3.75$\cdot 10^{-2}$& 2.35$\cdot 10^{-1}$& -1.18$\cdot 10^{-2}$& -3.99$\cdot 10^{-1}$& 2.18$\cdot 10^{-1}$\\
 \hline
$\iota$ &9.97$\cdot 10^{-3}$& 3.57$\cdot 10^{-1}$& -6.66$\cdot 10^{-3}$& -3.42$\cdot 10^{-1}$& 4.94$\cdot 10^{-1}$\\
\hline
\end{tabular}
\caption{Same as Table \ref{Table.SummaryBNS}, but using the BBH catalog.}\label{Table.SummaryBBH}
\end{table}

\begin{table}[htb]
\begin{tabular}{|c||c|c|c|c|c|}
\hline
 &$\overline{\Sigma}$&$\Delta\Sigma$&$\underline{\Sigma}$&$5^{th}$&$95^{th}$\\ 
 \hline
$\mathcal{M}$ &7.68$\cdot 10^{-3}$&1.02$\cdot 10^{-1}$&6.44$\cdot 10^{-3}$&-1.33$\cdot 10^{-1}$&1.51$\cdot 10^{-1}$\\
\hline
$\eta$ &7.27$\cdot 10^{-3}$&1.28$\cdot 10^{-1}$&8.84$\cdot 10^{-3}$&-1.45$\cdot 10^{-1}$&1.59$\cdot 10^{-1}$\\
\hline
RA & 1.44$\cdot 10^{-2}$&3.87$\cdot 10^{-1}$&8.35$\cdot 10^{-3}$&-4.25$\cdot 10^{-1}$&4.58$\cdot 10^{-1}$\\
\hline
dec.&-5.10$\cdot 10^{-2}$&4.49$\cdot 10^{-1}$&-2.43$\cdot 10^{-2}$&-5.32$\cdot 10^{-1}$&4.34$\cdot 10^{-1}$\\
\hline
$\psi$ &-3.68$\cdot 10^{-3}$&5.26$\cdot 10^{-2}$&-2.06$\cdot 10^{-3}$&-5.51$\cdot 10^{-2}$&5.09$\cdot 10^{-2}$\\
\hline
$\phi_0$ &-1.07$\cdot 10^{-3}$&5.16$\cdot 10^{-2}$&-1.72$\cdot 10^{-4}$&-5.37$\cdot 10^{-2}$&5.35$\cdot 10^{-2}$\\
 \hline
$t_0$ &-2.28$\cdot 10^{-2}$&4.05$\cdot 10^{-1}$&-5.88$\cdot 10^{-3}$&-3.33$\cdot 10^{-1}$&2.96$\cdot 10^{-1}$\\
 \hline
$D$ &-5.32$\cdot 10^{-2}$&2.80$\cdot 10^{-1}$&-1.72$\cdot 10^{-2}$&-5.14$\cdot 10^{-1}$&2.43$\cdot 10^{-1}$\\
 \hline
$\iota$&-8.15$\cdot 10^{-4}$&3.75$\cdot 10^{-1}$&-7.51$\cdot 10^{-3}$&-4.46$\cdot 10^{-1}$&4.92$\cdot 10^{-1}$\\
 \hline
\end{tabular}
\caption{Same as Table \ref{Table.SummaryBNS}, but using the BHNS catalog}\label{Table.SummaryBHNS}
\end{table}

The $\Sigma$s have means very close to zero for all the parameters, indicating that, when averaging over many events and the many CEs realizations, there are no preferred directions for CE-induced systematic biases in parameter estimates.
When it comes to the widths of the $\Sigma$ distributions, we can group the parameters into three different sets: 
\begin{itemize}
\item For the intrinsic parameters $\eta$ and $\mathcal{M}$, and the distance, the width is of the order $1-2 \times \sim10^{-1}$.
\item For the arrival time, the position parameters RA and dec, and the inclination, the widths are a few times larger, $\sim 3-5 \times 10^{-1}$.
\item The polarization and arrival phase have very large standard deviations, so the much smaller spread in their $\sigma$ is a consequence of their large standard deviations.
\end{itemize} 

\begin{figure}[htb]
\includegraphics[scale=0.35]{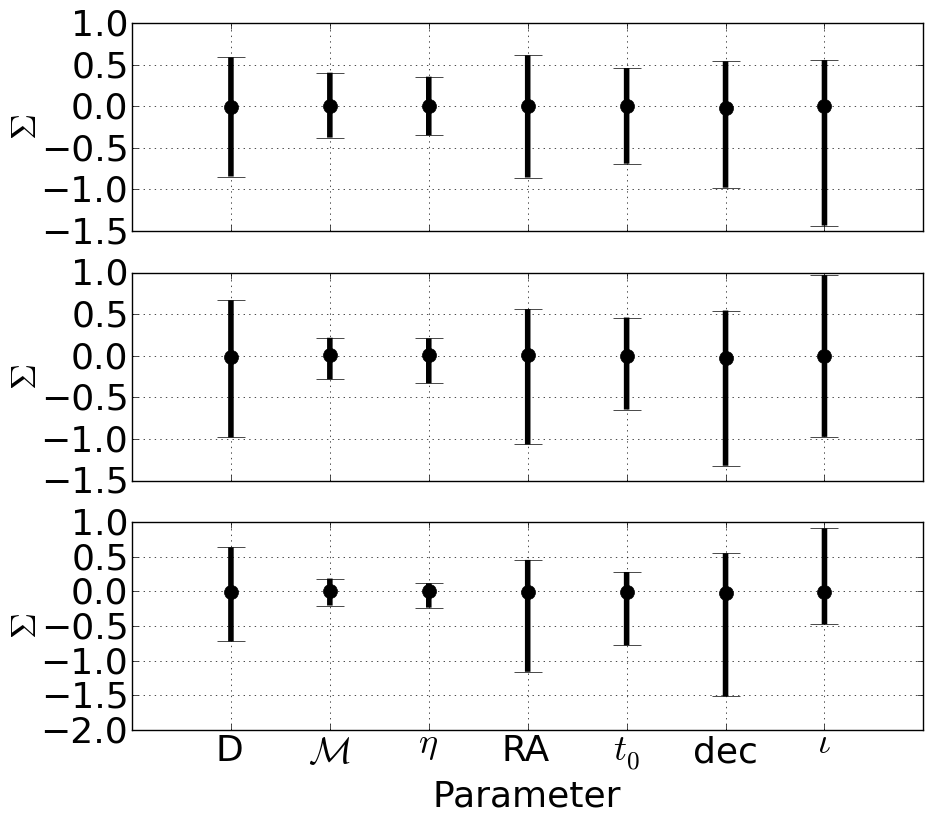}
\caption{The median of $\Sigma$ averaged among the 10 CE realizations. The lower end of the error bars corresponds to the lowest 5th percentile encountered in the various CE runs, while the upper end corresponds to the highest 95th percentile. We do not show $\psi$ and $\phi_0$ as those parameters are very poorly estimated. The upper panel refers to the BNS catalog, the middle one to the BHNS and the bottom one to the BBH catalog.}\label{Fig.MinMax_BNS}
\end{figure}

The averaged numbers we gave in Tables \ref{Table.SummaryBNS}, \ref{Table.SummaryBBH} and \ref{Table.SummaryBHNS} describe the \emph{typical} scenario, as they were obtained averaging among the 10 CE curves, reducing the impact of CE curves which had produced the largest spreads.  An alternative representation is shown in Fig.~\ref{Fig.MinMax_BNS}, where we plot the median of $\Sigma$ for each parameter (except $\psi$ and $\phi_0$, as we have seen they are always estimated with huge errors) averaged over the 10 CE realizations, with error bars whose min and max values are the \emph{worst} 5th and 95th percentiles encountered in the 10 CE runs.  These error bars yield a \emph{conservative} estimate of the impact of calibration errors when the actual CE realization and the statistics of the injection parameters line up to produce the largest shifts in parameter estimation.

Apart from the 1D results we have reported, it is interesting to verify how the confidence in our knowledge of the position of the source in the sky changes because of the CEs, as this will capture the \emph{joint} variation of RA and dec, taking into account their correlation. Let us call $\mbox{M}_e=(\mbox{dec}_e,\mbox{RA}_e)$ the point in the unit sphere whose spherical coordinates are given by the median value of RA and dec calculated in the exact run. Using the line element of a 2D sphere, we can write the size of the random error in the estimation of $\mbox{M}_e$ as $$\epsilon_e^2\equiv \Delta \mbox{dec}_e^2 + \sin\left(\frac{\pi}{2}-\mbox{dec}_e\right)^2 \Delta \mbox{RA}_e^2.$$ Adding the CEs will similarly yield the median sky location $\mbox{M}_m\mbox{=(dec}_m,\mbox{RA}_m)$, and we can measure the distance in the unit sphere between the points M$_e$ and M$_m$:

$$\epsilon_{me}^2= (\mbox{dec}_m - \mbox{dec}_e)^2  + \sin\left(\frac{\pi}{2}-\mbox{dec}_e\right)^2 (\mbox{RA}_m-\mbox{RA}_e)^2   $$

\begin{figure}[htb]
\includegraphics[scale=0.35]{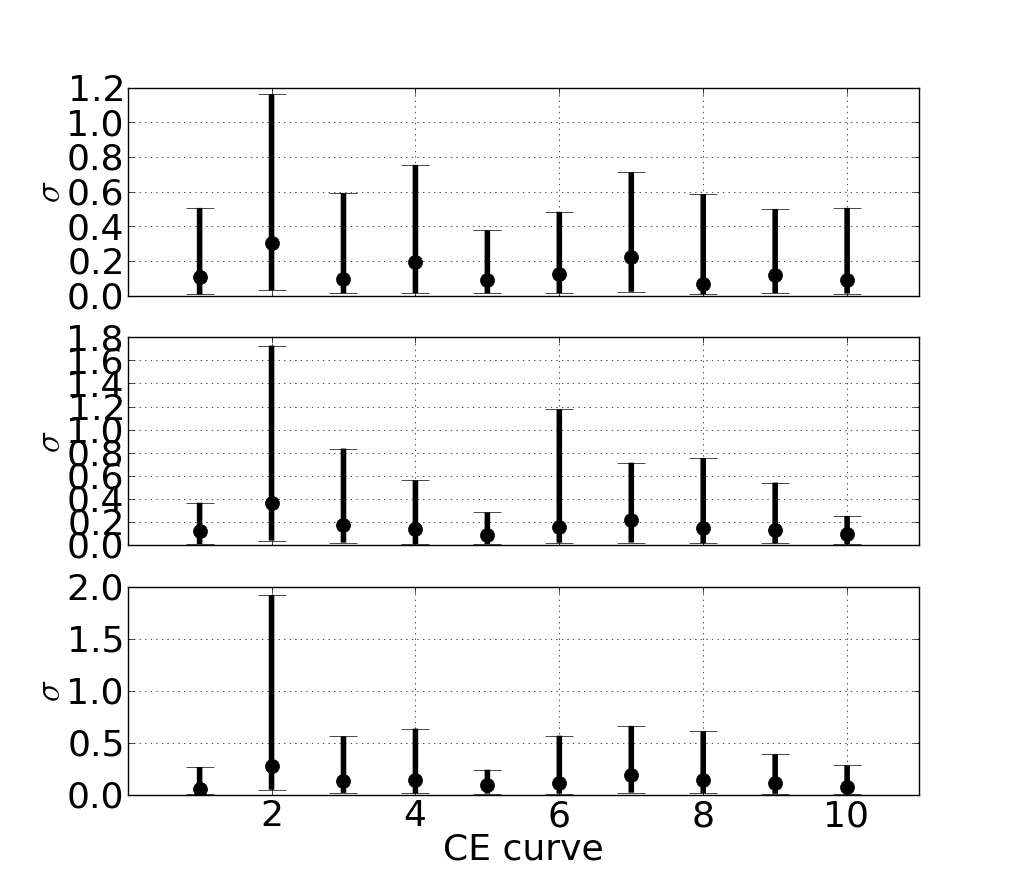}
\caption{The median of $\sigma$ (introduced in the main text) when using the various CE curves (shown in the abscissa label) and the three mass bins (from the top to the bottom: BNS, BHNS, BBH). The error bars show the 5th and 95th percentiles. Note that the ordinate scale varies in the subplots.}\label{Fig.SigmaCompare}
\end{figure}

We weight the distance between the exact and measured position in the unit sphere by the size of the random error box of the exact run:

\beq
\sigma\equiv \frac{\epsilon_{me}}{\epsilon_e},
\eeq
with $\sigma=0$ implying that the shift introduced by the CEs is null, and $\sigma>1$ that it is larger than the uncertainties due to the noise.
In Fig.~\ref{Fig.SigmaCompare} we show the median of $\sigma$, together with 5th and 95th percentiles, for all the CEs and the three mass bins.

It is evident that CE curve 2 leads to average shifts which are much larger than for the other CE curves (the median of $\sigma$ is larger than $0.5$ in the three catalogs), and to very large spreads (95th percentile larger than $1.6$). 
Note however that we are weighting the distance in the unit sphere by the width of the random error box of the exact run. Thus a large value of $\sigma$ does not imply a large shift in \emph{radians}. We have indeed verified that some of the signals that go in the tails of the distribution of $\sigma$ are high-SNR signals, for which $\epsilon_e$ is very small, and so is $\epsilon_{me}$, even though their ratio may be $\sim 2-3$.
  
It is known that in a three-interferometer network, if the position of the source were to be estimated using just time triangulation, there would be a degeneracy corresponding to a reflection of the position with respect to the plane that contains the three IFOs \cite{Fairhurst2010,JaranowskiAl1996}. In reality, amplitude information and correlations with the remaining parameters also affect the sky localization (e.g. disentanglement of the plus and cross polarization) and break this symmetry. In this way, one of the two specular positions can be actually preferred and assigned a higher probability.  Perturbations to the phase of the injected signal, like the ones introduced by the calibration phase errors, may change the situation and push our inference towards the reflected position.

We have found three signals (one in the BBH catalog and two in the BHNS catalog) for which adding the CEs leads to the aforementioned behavior. More precisely in two cases the signal was found in the specular position with respect to the IFOs plane; in the third case it was found in a position belonging to the ring on the sky which assures the same H1-L1 time delay (this is discussed for example in~\cite{Raymond2009} for a network made of H1 and L1 only. Although we are using three IFOs in this work, for the event we are discussing now, the SNR in Virgo was 4 times smaller than the SNR in H1 and L1, which explain why the result is similar to a H1-L1 network.).
This phenomenon happened only with a few CEs curves (3 out of 10). After a thorough analysis we have concluded that this behavior was not solely due to the addition of CEs but also to the particular noise realizations for those events. 
In fact, we have rerun the analysis on those signals, using 100 different noise realizations, finding that only $8\%$ of the noise streams, in conjunction with the CEs,  would lead to the aforementioned large shifts.
Considering that these outliers were nine (3 signals times 3 CEs curves), over the initial set of 7500 signals, and that only 8 noise realizations over 100 produced them, we concluded that the probability of such extreme shifts is $\sim 0.1\%$ and we did not take them into consideration while writing Tables \ref{Table.SummaryBBH} and \ref{Table.SummaryBHNS}.

\subsection{Effects on Bayes factors}

The main outcome of the Nested Sampling code is the Bayes' factor, a measure of the confidence in the hypothesis that a signal is buried in the noise. 

To be more precise, the evidence (Eq.~\ref{Evidence}), and thus the Bayes' factor, which is the ratio between the evidence of two models (Eq.~\ref{Odds}), is the measure of the fit of the data to the model. Being marginalized over all the parameters, it shows the \emph{mean} match between the model and the data. 
Because of its huge range of variation, it is usually the log of this quantity which is reported, logBSN. Hence we will quote the natural logarithm of the Bayes' factor, as defined in Eq.~\ref{Odds}. 

In~\cite{Veitch2010} a method was described in which the logBNS could be used as a detection statistic. It was shown how, if one assigned equal prior probability to the presence of a signal, as  opposed to the presence of pure noise, a threshold of $\mbox{BSN } \sim 2.8$ could be set, such that  the $99\%$ of the analyses which gave a BNS $ >2.8$ contained a signal. A more refined estimation, which takes into account our knowledge on the rates with which GWs should be detected, sets this threshold to $\sim 20$~\cite{Veitch2008}.

\begin{figure}[h]
\includegraphics[scale=0.35]{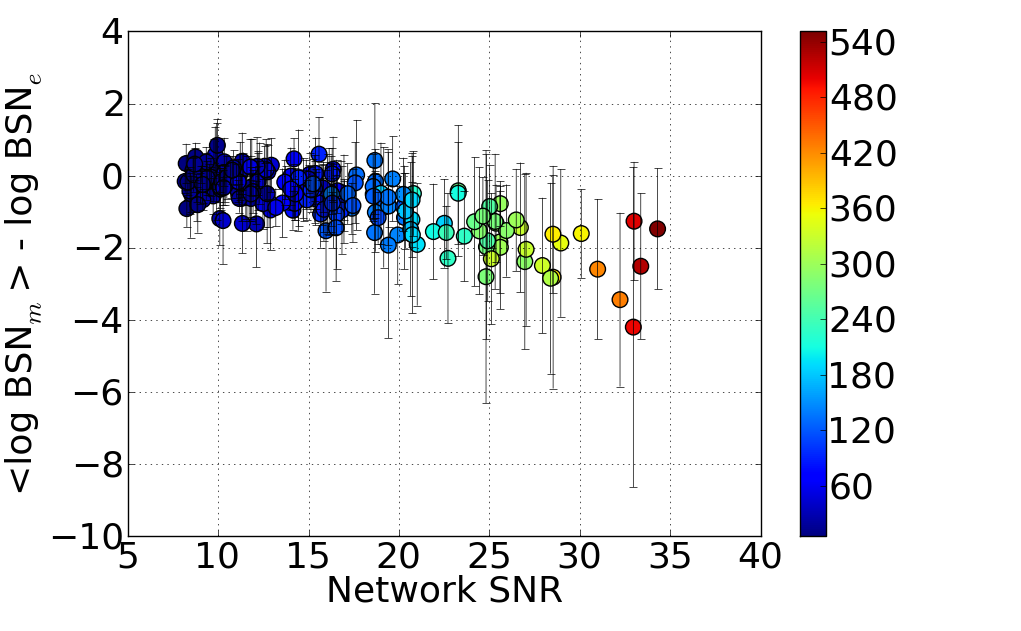}
\caption{(color online) The difference in log Bayes factor between the exact run and the average of the runs with calibration errors. The values in the colorbar correspond to the $\mbox{BSN}_e$ produced by the injections. Generally, louder signals are affected by larger shifts.}\label{Fig.BayesAverageDiff_BHNS}
\end{figure}

It is then interesting to study, beside the systematics that CEs introduced in the estimated parameters, the effects they might have in the estimation of the Bayes factor, as large shifts may decrease the confidence we assign to a detection. Moreover, comparing the bayes factors with and without calibration errors is a direct measure of how much worse the fit is overall. We have complemented our analysis by investigating this issue
In Fig.~\ref{Fig.BayesAverageDiff_BHNS}  we show, for all the injections in the BHNS catalog, the difference between the average of the measured logBNS over the ten CE realizations and the exact log Bayes factor, $\mbox{logBSN}_e$: \mbox{$ \langle \mbox{logBNS}_{m} \rangle-\mbox{logBSN}_e$}, where we have indicated with wedge brackets the average over the CE realizations:  $\langle \mbox{logBNS}_{m} \rangle\equiv \frac{1}{10} \sum_{i=1}^{10} \mbox{logBNS}_{m}^{(i)}$, plotted against the optimal SNR \footnote{As the optimal SNR is unaffected by CEs, Eq.~(\ref{OptimalSNRUnchanged}), we are allowed in Fig.~\ref{Fig.BayesAverageDiff_BHNS} to use a single $x$ axis.}. We also show error bars corresponding to the spread of $\mbox{logBSN}_m$ amongst the CE realizations and we colored the points according to the $\mbox{logBSN}_e$ of the injections. 

It is evident from Fig. \ref{Fig.BayesAverageDiff_BHNS} that the higher the optimal SNR (and consequently logBSN) of the injected signal, the bigger the impact of CEs on the logBSN. In fact, a signal with a high SNR will be ``clearly'' detected by the PE code, and well matched with the right template. In this scenario, the disturbances due to CEs are more visible (\emph{i.e.} the change in logBSN larger) than in a low SNR scenario. When the signal is hardly detected, CEs add only some extra mismatch.
In general the effects are very small, the average shift in logBSN over the three mass bins and the 10 CEs curves we have considered being $0.9\%$, with the binary neutron star systems being the most affected ($1.8\%$).

We can then conclude that, if the Bayes factor was used as a complimentary piece of data in assessing the confidence of a detection, it would represent a reliable help, being barely affected by calibration errors.

\subsection{Comparing the CE realizations}\label{Sec.Compare}

The data analyst will not know the exact shape and magnitude of the CE the data are being affected from; it is then an interesting exercise to study how the effects of the errors vary with the CE curves' shape. 

\begin{figure}[h]
\includegraphics[scale=\Figscale]{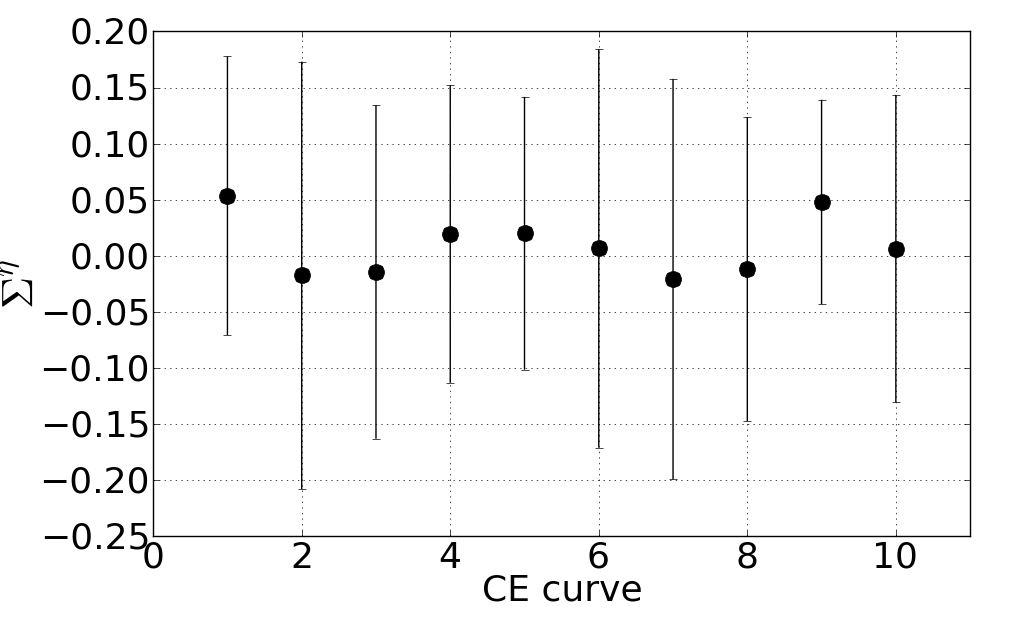}
\caption{Mean of $\Sigma^{\eta}$ with the various CE realizations for the BHNS catalog}\label{Fig.CompareEta}
\end{figure}

To study how parameter estimation reacts to the CE curves, we show how the median and standard deviations of the $\Sigma$s of the various parameters vary among the ten CE realizations in the three catalogs. For example, in Fig. \ref{Fig.CompareEta} we plot the median of $\Sigma^{\eta}$ (mass ratio) over the injections in the BHNS catalog, together with their standard deviations, for each CE realization (labeled in the X axis). 
\\

\begin{figure}[h]
\includegraphics[scale=\Figscale]{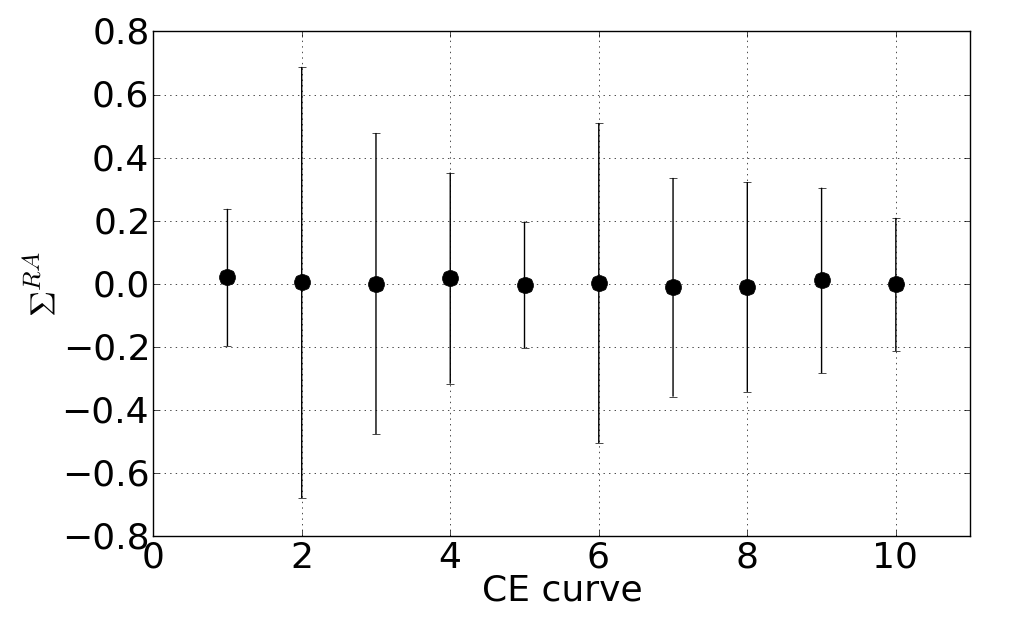}
\caption{Mean of $\Sigma^{RA}$ with the various CE realizations for the BHNS catalog}\label{Fig.CompareRa}
\end{figure}

It is quite remarkable as all the CEs give $\Sigma^{\eta}$ with similar averages, the largest difference being $\sim 0.07$. A similar plot is obtained for the chirp mass.
In Fig. \ref{Fig.CompareRa} we show the same plot for RA (note that the y axis scale is much larger than in Fig. \ref{Fig.CompareEta}). For RA and dec the results of the runs with the various CEs are comparable, but the error bars are generally larger than for the intrinsic parameters, meaning that those parameters are more affected by the calibration errors. Sky localization is most strongly affected by differences in amplitude calibration errors in different interferometers at frequencies where the interferometers are most sensitive. This is particularly true for Hanford and Livingston interferometers, which are relatively nearby and nearly aligned, meaning that any incoherence in the recovered amplitudes can not be fit by adjusting the inclination or polarization of the source, and can influence the recovered sky location. Therefore, it is not surprising to see much larger variations for the second and sixth CE realizations, for which the amplitude corrections for H1 and L1 have opposite signs near $100$ Hz (see Figs. \ref{Fig.Errors2} and \ref{Fig.Errors6}).  Meanwhile, e.g., the fifth CE realization has very comparable amplitude CEs for H1 and L1 at $100$ Hz (see Fig. \ref{Fig.Errors5}), matching up to the small range of normalized systematic biases in RA (see Fig.~\ref{Fig.CompareEta}).

\begin{figure}[h]
\includegraphics[scale=\Figscale]{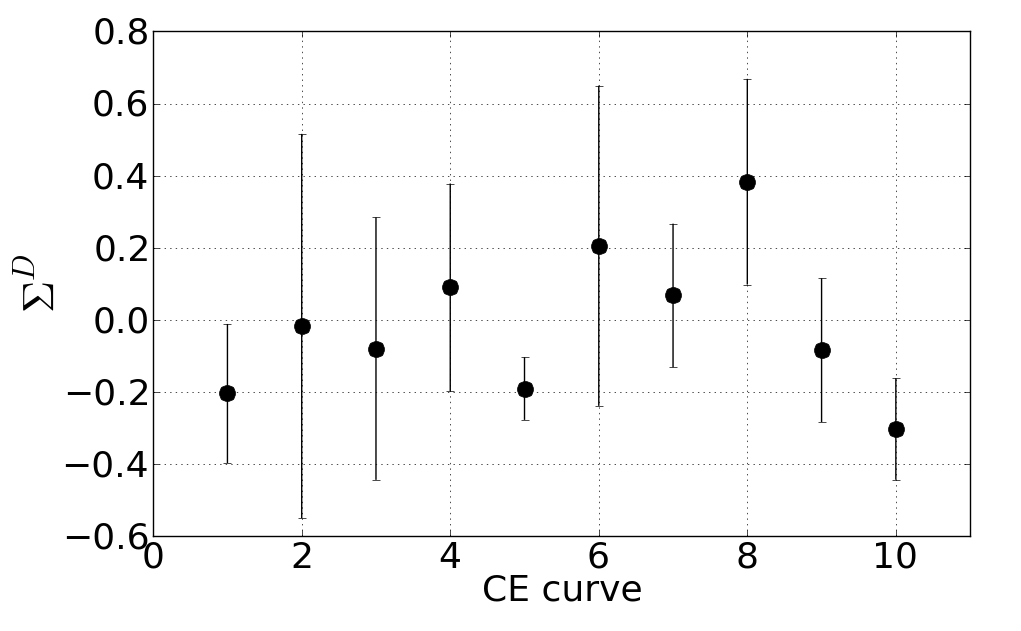}
\caption{Mean of $\Sigma^{D}$ with the various CE realizations for the BHNS catalog. }\label{Fig.CompareDist}
\end{figure}

The medians of $\Sigma$ for the distance, Fig. \ref{Fig.CompareDist}, are not centered around zero. This is not unexpected, as we have pointed out earlier that the distance estimation is directly affected by the amplitude errors.

\section{Conclusions}

In this work we have quantified in a systematic way, for the first time in the literature, the effects of calibration errors on the estimation of parameters of gravitational waves emitted by binary systems with non-spinning components. 
We have considered three mass bins, and for each bin we have created a catalog with 250 sources, uniformly distributed in the sky. 
A Bayesian parameter estimation code was run on all the injections of these catalogs, first using the exact transfer function (i.e., without calibration errors), and then after transforming the data with one of the ten calibration error curves we have generated. We have then compared the posterior distributions, as well as the Bayes factors, of the runs where the errors were added with the control runs, where no errors were present. 

We found that for all the error curves considered, the effects are small, the systematic shift introduced in the estimated parameters being a fraction of the statistical measurement errors. 
We also considered the effect of calibration errors on Bayes factors, finding that it is larger for louder injections, but always small enough that no signals would be missed because of calibration errors by a putative pipeline that would rank events by Bayes factors.

Furthermore, we have found that the different calibration error curves we considered yield compatible results, implying that the distribution of CE-induced shifts in parameter estimates does not strongly depend on the exact shape of the CE curves.

The inclusion of spins in the waveform model will lead to additional complications, and should be the subject of a future investigation.

\section{Acknowledgements}

The authors would like to thank Will Farr for useful comments on the first draft of this paper, Duncan Brown, Gabriela Gonzalez, Romain Gouaty, Keita Kawabe, Nobuyuki Kanda, Giulio Mazzolo, Beno\^{i}t Mours, David Rabeling, Lo\"{i}c Rolland,  Francesco Salemi, Xavier Siemens, Alberto Vecchio, Alan Weinstein, Michele Zanolin and the Bayesian group for useful discussions.\\
The authors acknowledge the use of computing resources: the Atlas cluster at the Albert Einstein Institute in Hannover, supported by the Max Planck Gesellschaft, and the Nemo cluster supported by National Science Foundation awards PHY-0923409 and PHY-0600953 to UW-Milwaukee.\\
SV, WDP, TGFL and CVDB are supported in part by the research program of the Foundation for Fundamental Research on Matter (FOM), which is partially supported by the Netherlands Organisation for Scientific Research (NWO).\\
BA is funded by a Science and Technology Facilities Council studentship.\\
JV acknowledges the STFC Grant No. ST/J000345/1

\appendix
\section{Error curves}\label{App.CECurves}

In this section we show nine of the ten calibration error curves. The remaining one was given in the main text, Fig. \ref{Fig.Errors1}.

\begin{figure}[htb]
\includegraphics[scale=\Figscale]{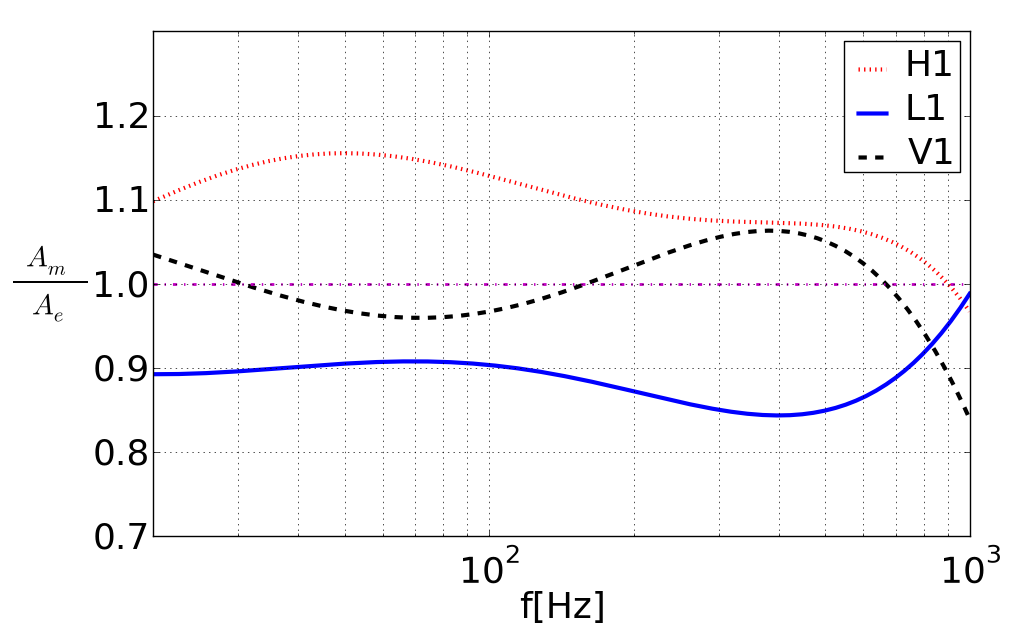}\\
\includegraphics[scale=\Figscale]{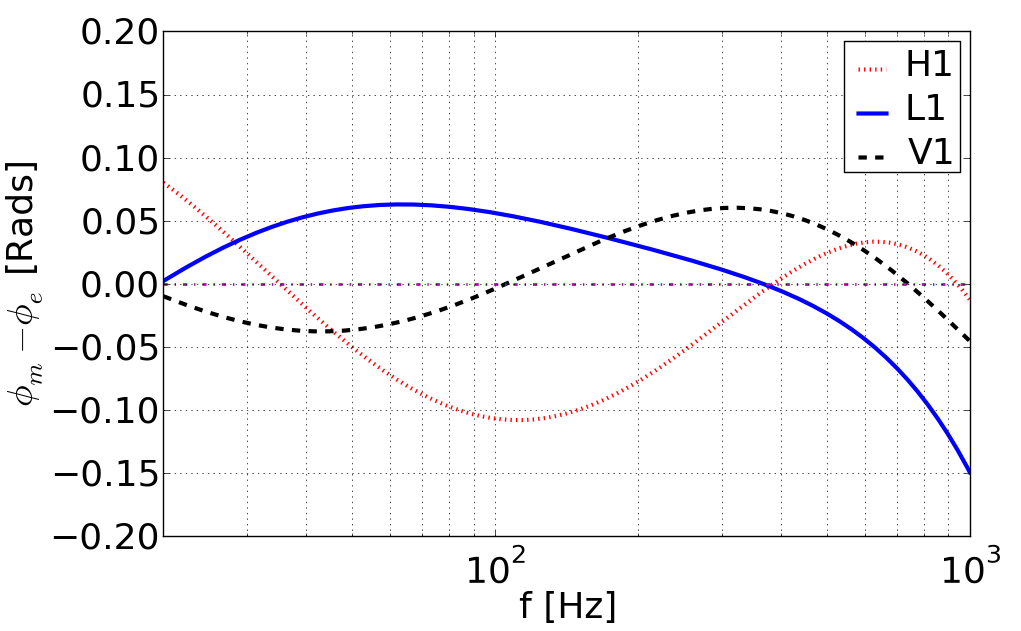}
\caption{The second CE realization for the amplitude (top) and phase (bottom).}\label{Fig.Errors2}
\end{figure}
\begin{figure}[htb]
\includegraphics[scale=\Figscale]{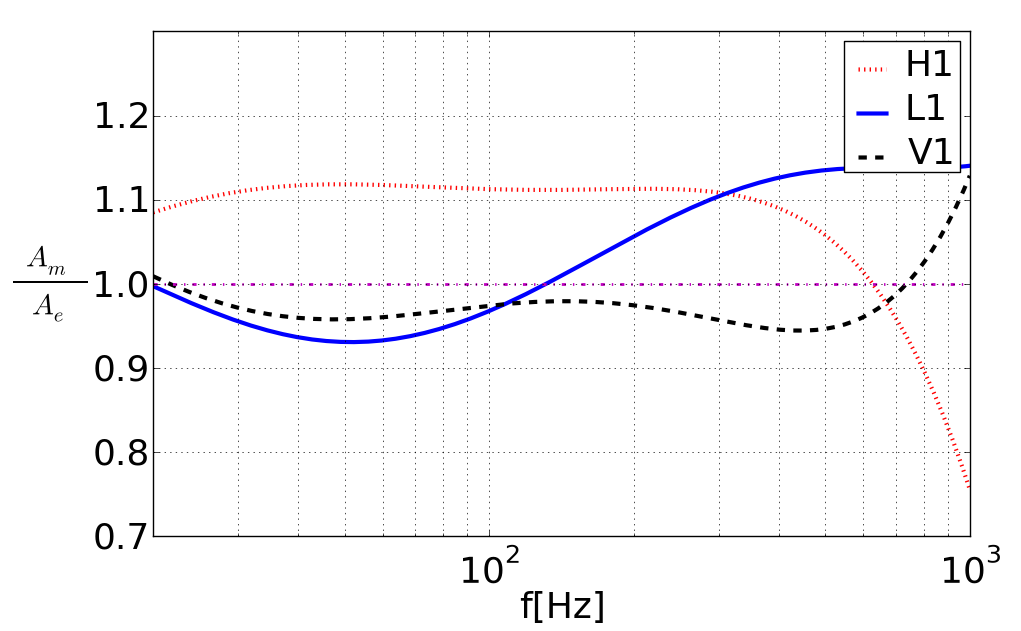}\\
\includegraphics[scale=\Figscale]{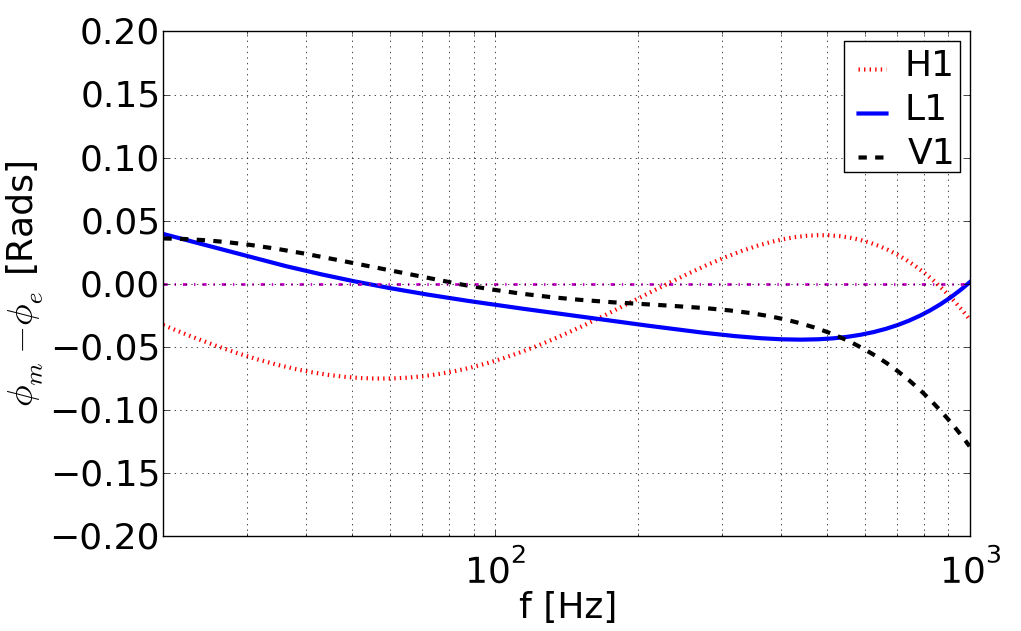}
\caption{The third CE realization for the amplitude (top) and phase (bottom).}\label{Fig.Errors3}
\end{figure}
\begin{figure}[htb]
\includegraphics[scale=\Figscale]{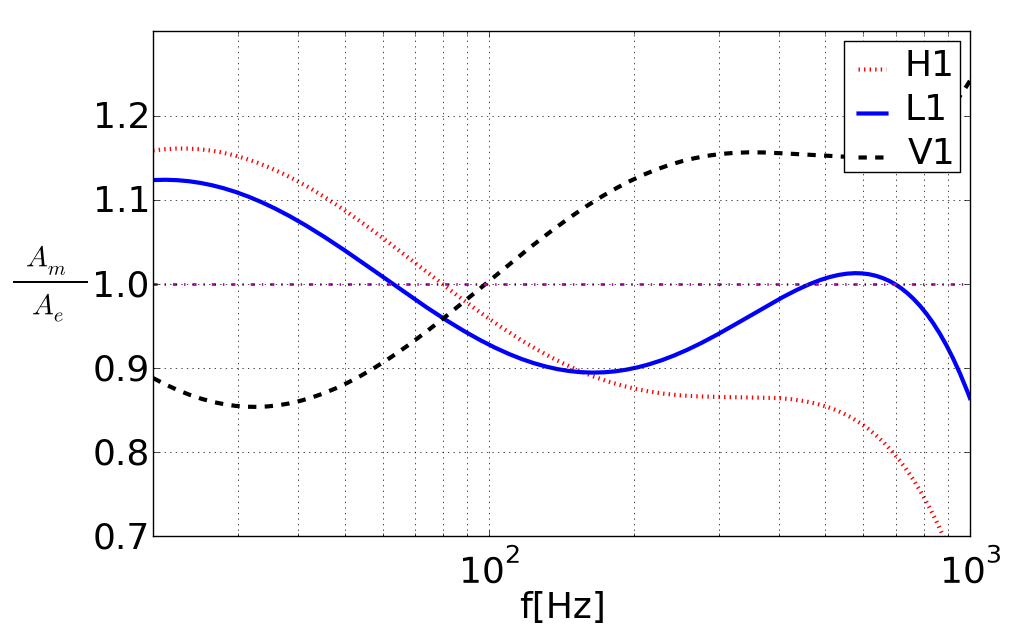}\\
\includegraphics[scale=\Figscale]{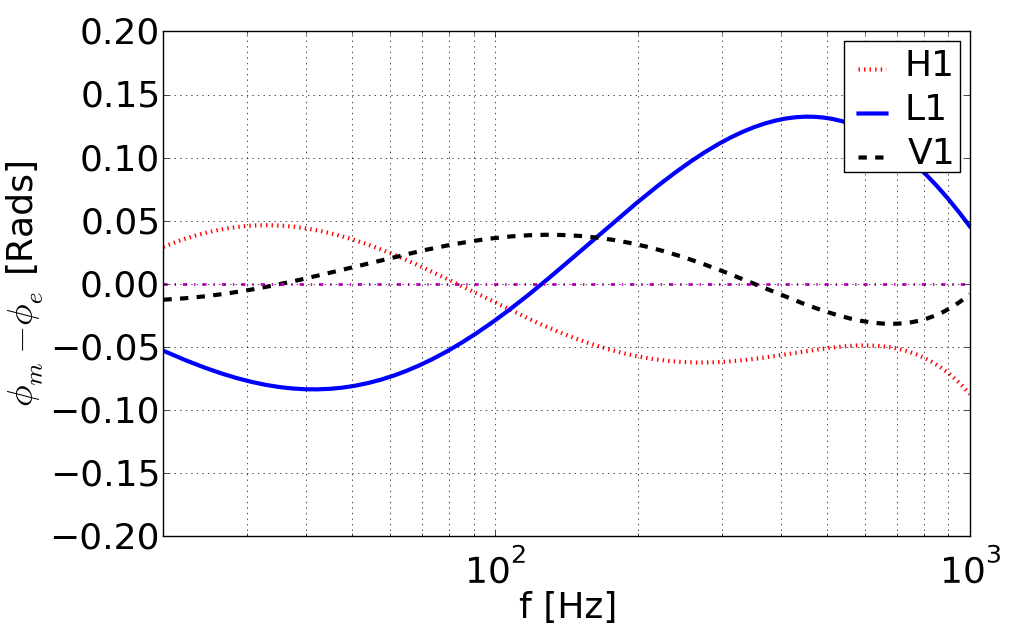}
\caption{The fourth CE realization for the amplitude (top) and phase (bottom).}\label{Fig.Errors4}
\end{figure}
\begin{figure}[htb]
\includegraphics[scale=\Figscale]{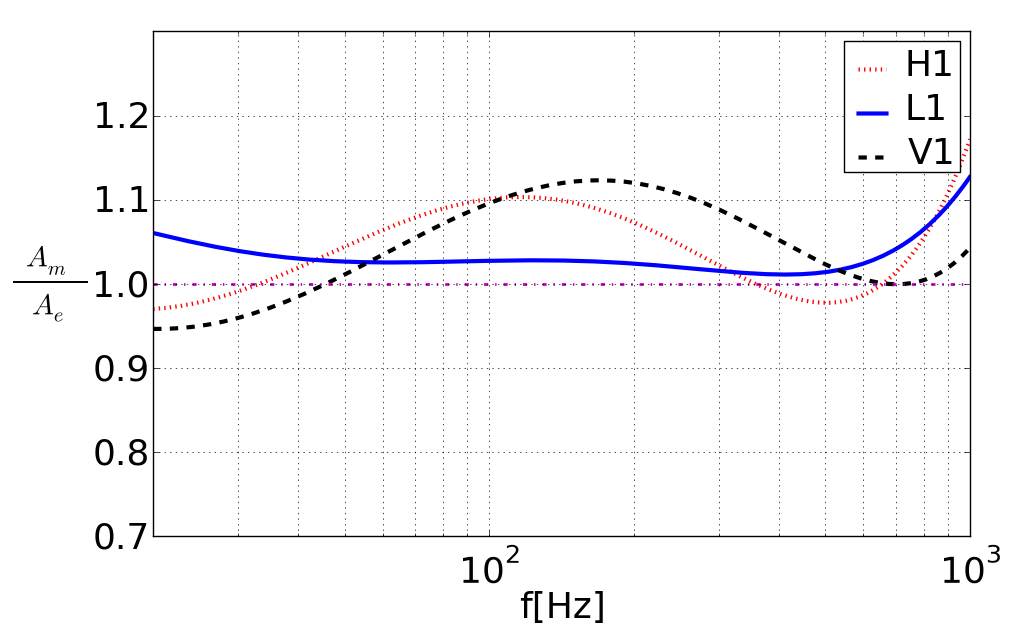}\\
\includegraphics[scale=\Figscale]{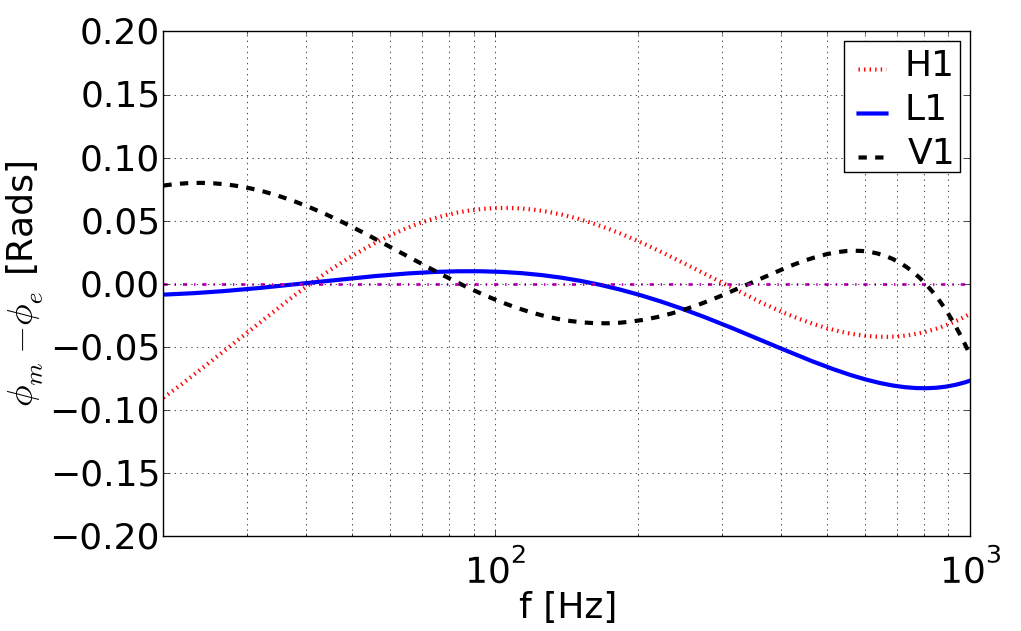}
\caption{The fifth CE realization for the amplitude (top) and phase (bottom).}\label{Fig.Errors5}
\end{figure}
\begin{figure}[htb]
\includegraphics[scale=\Figscale]{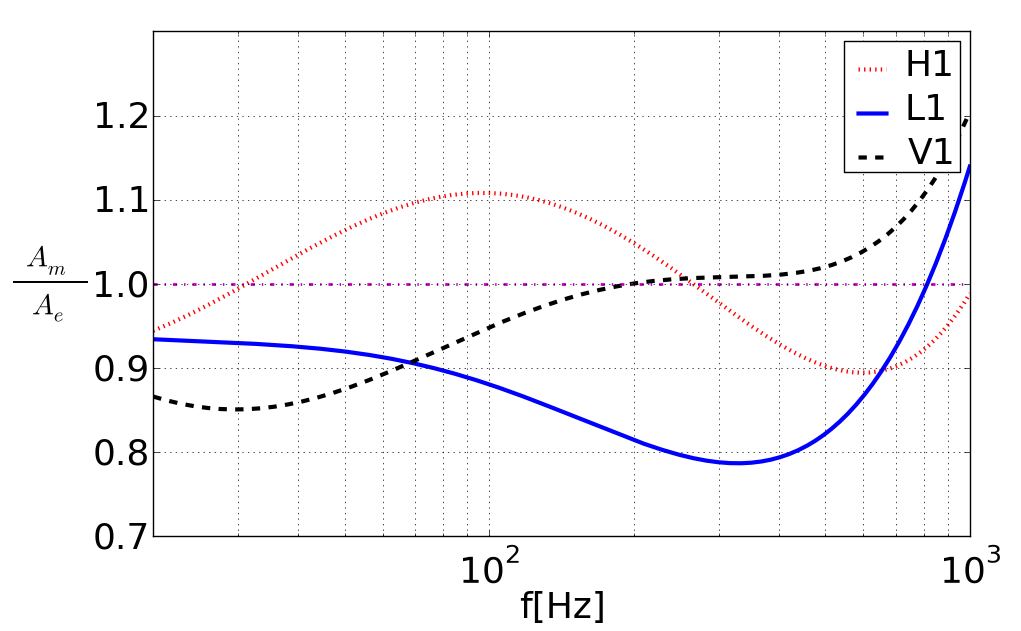}\\
\includegraphics[scale=\Figscale]{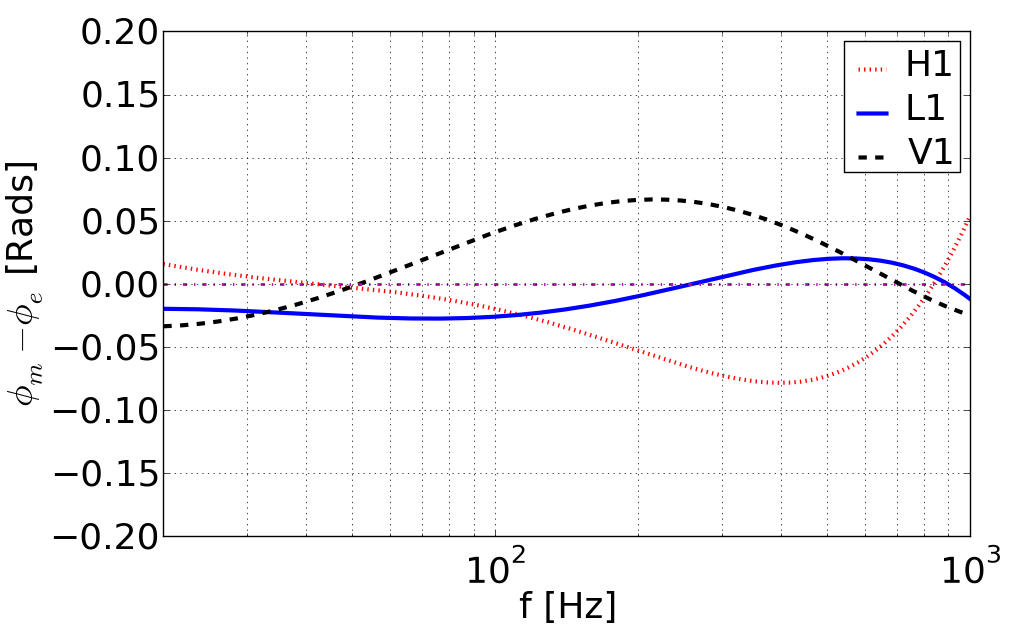}
\caption{The sixth CE realization for the amplitude (top) and phase (bottom).}\label{Fig.Errors6}
\end{figure}
\begin{figure}[htb]
\includegraphics[scale=\Figscale]{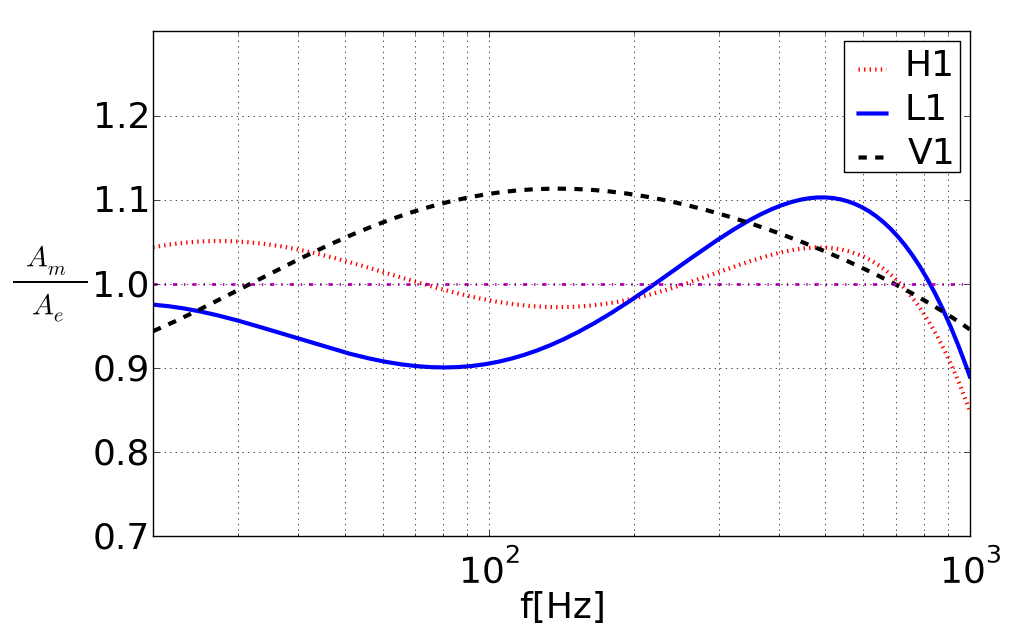}\\
\includegraphics[scale=\Figscale]{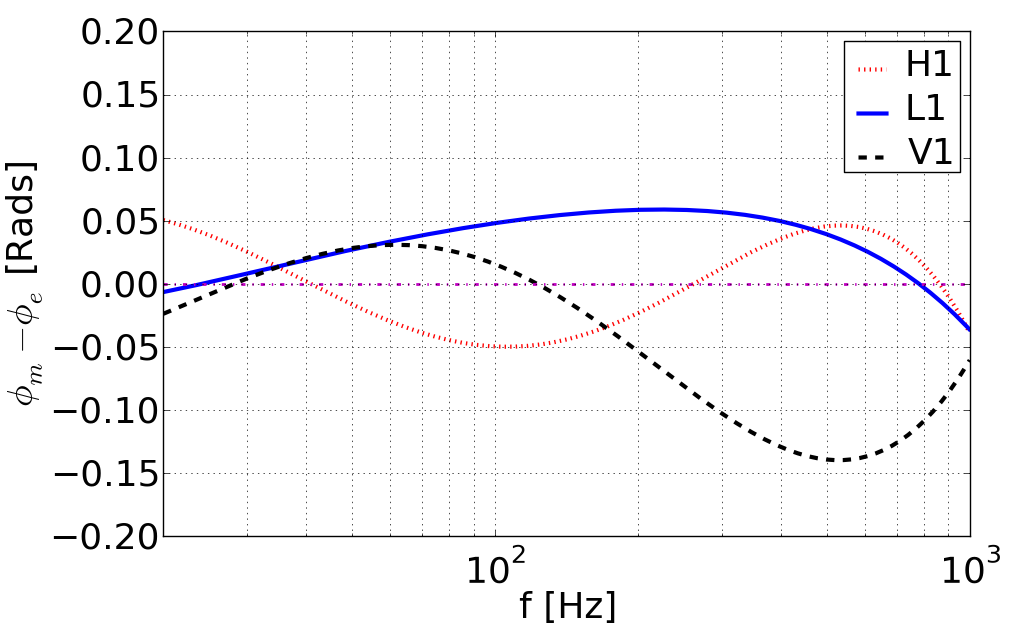}
\caption{The seventh CE realization for the amplitude (top) and phase (bottom).}\label{Fig.Errors7}
\end{figure}
\begin{figure}[htb]
\includegraphics[scale=\Figscale]{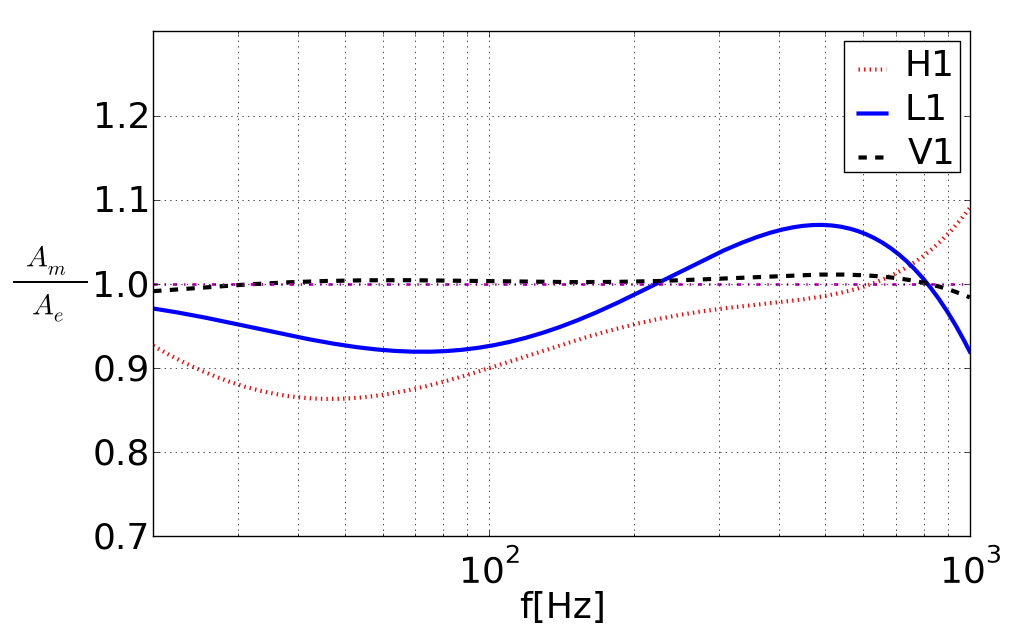}\\
\includegraphics[scale=\Figscale]{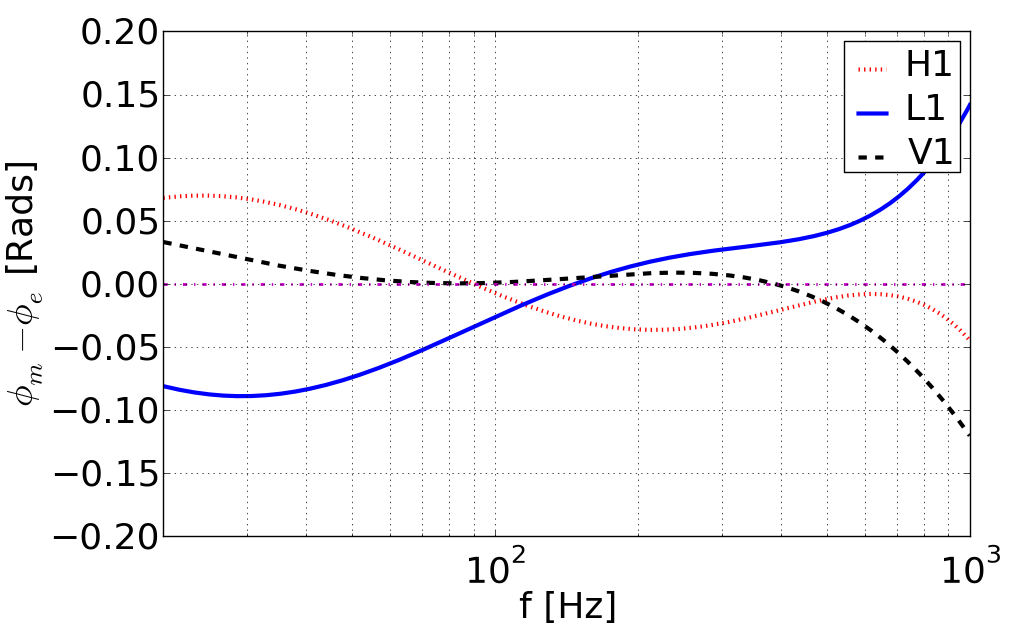}
\caption{The height CE realization for the amplitude (top) and phase (bottom).}\label{Fig.Errors8}
\end{figure}
\begin{figure}[htb]
\includegraphics[scale=\Figscale]{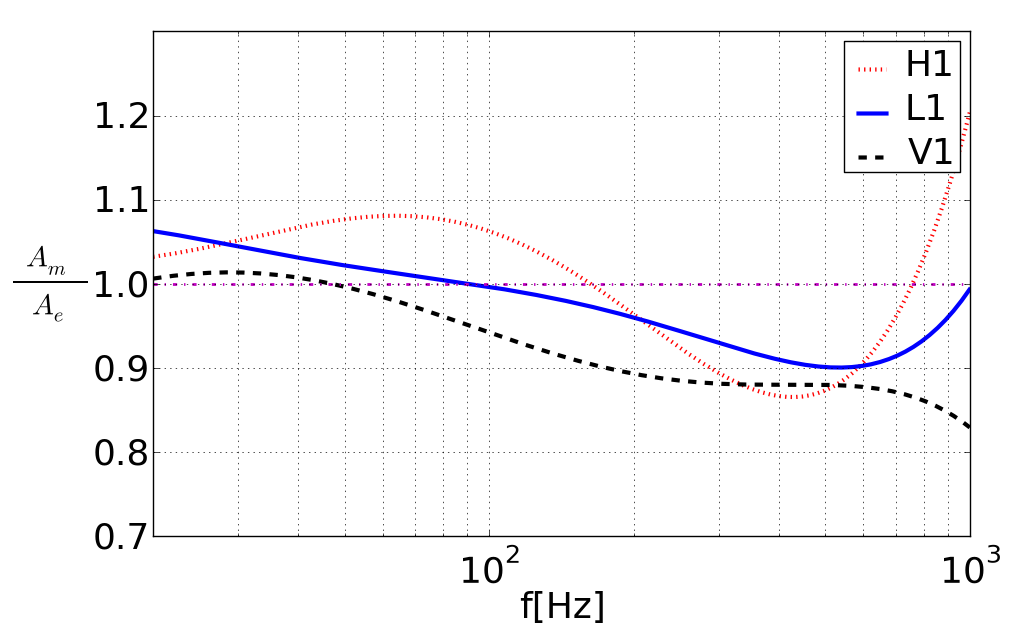}\\
\includegraphics[scale=\Figscale]{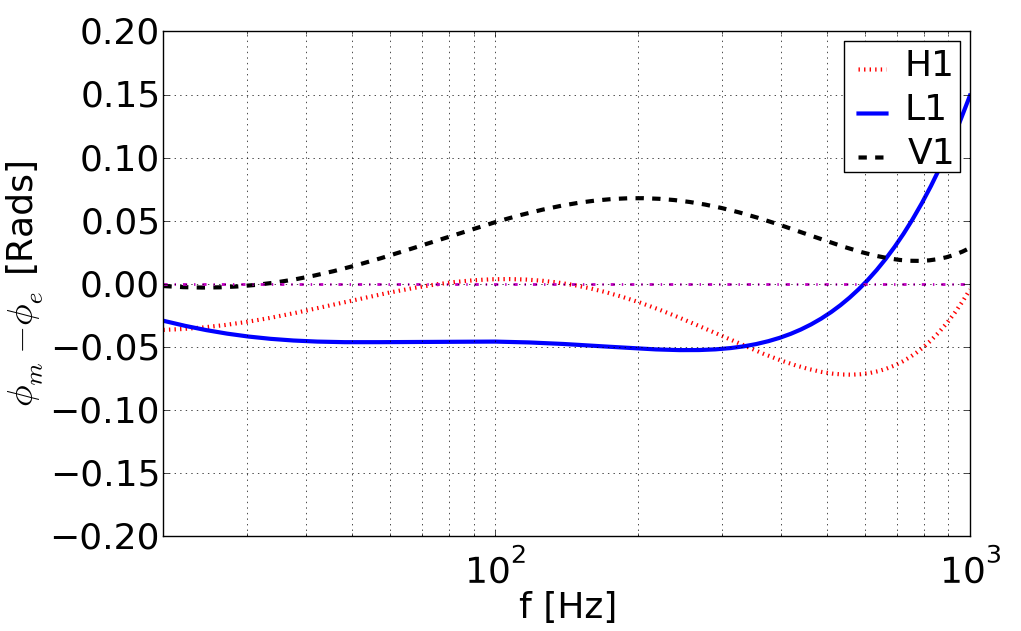}
\caption{The ninth CE realization for the amplitude (top) and phase (bottom).}\label{Fig.Errors9}
\end{figure}
\begin{figure}[htb]
\includegraphics[scale=\Figscale]{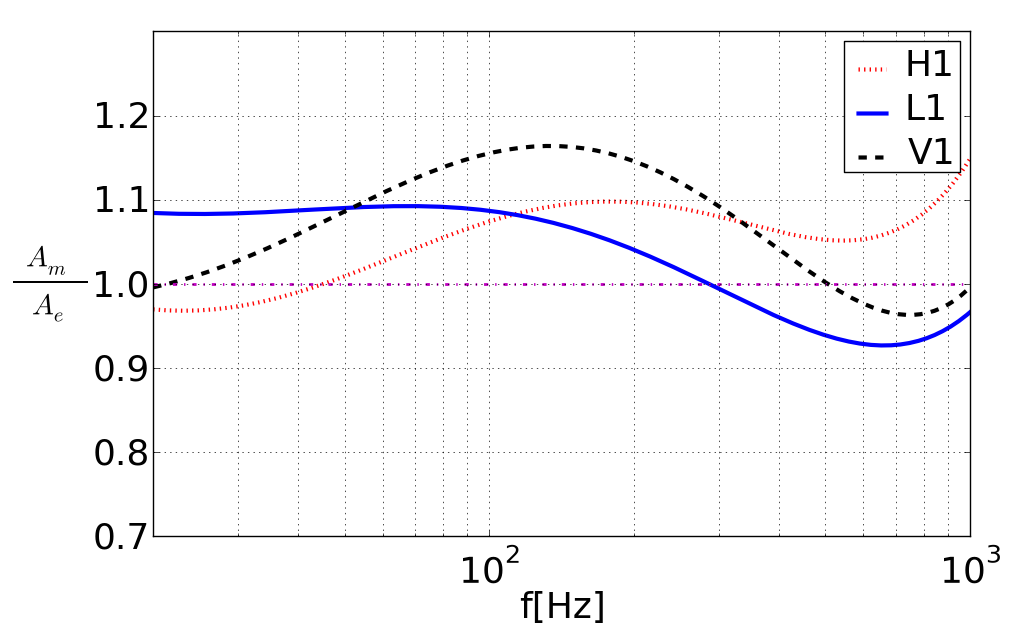}\\
\includegraphics[scale=\Figscale]{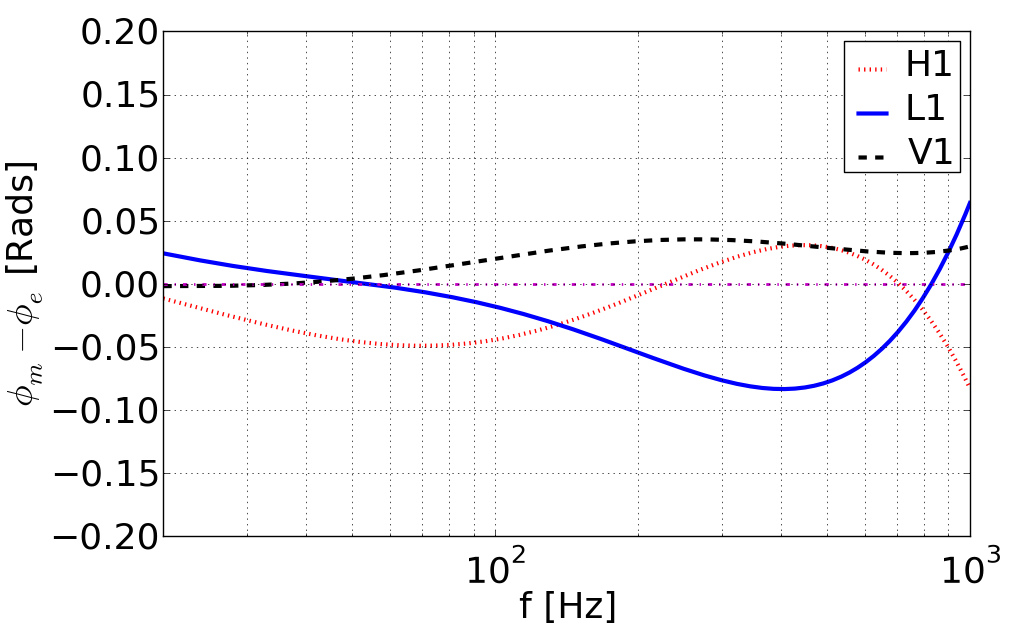}
\caption{The tenth CE realization for the amplitude (top) and phase (bottom).}\label{Fig.Errors10}
\end{figure}

\end{document}